\documentclass[a4paper,twocolumn,showpacs,superscriptaddress,floatfix]{revtex4-2}
\usepackage{graphicx}
\usepackage{epsf}
\usepackage{epsfig}
\usepackage{amssymb,amsmath}
\usepackage[usenames]{color}
\usepackage{amssymb}
\usepackage{times}
\usepackage{amsmath}
\usepackage{float}
\usepackage{mathtools}
\usepackage{mathrsfs}   
\usepackage{enumerate}

\newcommand{\be}{\begin{equation}}
\newcommand{\ee}{\end{equation}}
\newcommand{\bea}{\begin{eqnarray}}
\newcommand{\eea}{\end{eqnarray}}
\newcommand{\nn}{\nonumber}

\font\tenscr=rsfs10 scaled1100
\font\sevenscr=rsfs7 
\font\fivescr=rsfs5 
\skewchar\tenscr='177
\skewchar\sevenscr='177
\skewchar\fivescr='177
\newfam\scrfam
\textfont\scrfam=\tenscr
\scriptfont\scrfam=\sevenscr
\scriptscriptfont\scrfam=\fivescr

\begin{document}

\title{A geometric approach to QNMs in optics: application to pseudospectrum and structural stability}



\author{Lamis Al Sheikh}
\affiliation{School of Electrical and Electronics Engineering, Nanyang Technological University, 639798 Singapore}
\affiliation{Jülich Centre for Neutron Science (JCNS), Forschungszentrum J$\ddot{u}$lich, Garching, Germany}
\affiliation{Institut de Math\'ematiques de Bourgogne UMR 5584,
  Universit\'e Bourgogne Europe, CNRS, F-21000 Dijon, France}
\author{Jos\'e Luis Jaramillo}
\affiliation{Institut de Math\'ematiques de Bourgogne UMR 5584,
  Universit\'e Bourgogne Europe, CNRS, F-21000 Dijon, France}
\begin{abstract}
	We develop a geometric--spectral framework for the computation and stability analysis of quasi-normal modes (QNMs) in open optical cavities of compact support. The hyperboloidal approach, transferred from gravitational physics to electromagnetism \cite{Jaramillo:2020tuu}, incorporates outgoing boundary conditions directly into the formulation of the optical resonance problem. Dispersive and absorbing media are described by a Lorentz-model permittivity through an auxiliary-field formulation, and the resulting non-Hermitian spectral problem is solved using Chebyshev spectral discretization.Pseudospectrum analysis is then extended to the optical setting following \cite{Alsheikh21} and used to study the stability of optical resonances under perturbations. The pseudospectrum provides information on the sensitivity of the resonances that is not contained in the QNM spectrum alone. Particular attention is given to the role of the norm used to define the pseudospectrum, by comparing the stability properties obtained with different choices of scalar product. The results show that the assessment of spectral stability is closely related to the functional setting in which perturbations are measured. Combining the hyperboloidal formulation with pseudospectrum analysis therefore makes it possible to compute optical QNMs while also studying their spectral stability. The approach extends techniques developed for resonance problems in gravitational physics to dispersive electromagnetism and provides a basis for studying stability and sensitivity in open optical systems.
\end{abstract}

\pacs{}

\maketitle
\section{Quasi-normal modes as a non-selfadjoint spectral problem}
Quasi-normal modes (QNMs) are systematically used to describe resonances in open systems. In optical systems of compact support, the physical interfaces of the cavity satisfy the usual electromagnetic interface conditions, allowing reflection and transmission between the cavity and its exterior. When considering the QNMs, or natural modes, of the system, no incoming excitation is prescribed: the modal solutions are obtained in the absence of waves incident on the cavity. This should be distinguished from the outgoing boundary condition imposed at (null) infinity. Here, "outgoingness" is an asymptotic condition on the wave field: radiation is allowed to leave the physical domain through infinity, whereas no radiation enters the domain from infinity. This is the notion of outgoingness encountered in gravitational-wave and other open wave problems, and is conventionally expressed through a Sommerfeld-type radiation condition. Crucially, the study of QNMs in 
(photonics/plasmonics) optical systems,  has experienced
an extraordinary development in recent years
\cite{sauvan2013theory,Muljarov:18,LalYanVyn17}.

This work introduces two contributions to the spectral analysis of QNMs in open optical systems. First, we adapt the hyperboloidal compactification technique—originally developed in gravitational physics—to optical cavities. This approach yields a geometrically natural and mathematically-sound treatment of outgoing boundary conditions, avoiding the need for effective layers or absorbing boundaries (note, however, that these
schemes can be harmonised from a structural point of view, cf. \cite{zenginouglu2026null,wess2026finite}).
This works present the details and extend the discussion in  \cite{Alsheikh21} that represents, to our
knowledge,
the first application of the hyperboloidal compactification approach to the modeling of optical resonators (see also
\cite{10.3389/fphy.2024.1457543} for a recent contribution along these lines,  
extending the hyperboloidal scheme to the dispersive case with a focus on the study of dispersive extended theories of gravity).

Second, we incorporate pseudospectrum analysis into the study of electromagnetic QNMs. While pseudospectra have been crucial to understand non-normal operators in jl fluid dynamics \cite{TreTreRed93,trefethen2005spectra}, quantum mechanics \cite{KreSieTat15},  gravitational physics \cite{Jaramillo:2020tuu}, condensed matter and applied 
mathematics \cite{ColRomHan19} and other subdisciplines
 (see \cite{BizGasJar26} for a recent discussion of the
interdisciplinarity of the pseudospectrum),
their role in electromagnetism has remained not fully explored or employed. This work suggests the use of pseudospectrum notions in optics, in a geometrical framework similar to that introduced in gravitational physics \cite{Jaramillo:2020tuu,Alsheikh21}. Pseudospectrum potentially gives information about spectral instability of the non-selfadjoint QNM operator governing open optical systems.

From a mathematical perspective, QNM frequencies—also termed scattering resonances—correspond to the poles of the meromorphic continuation of the resolvent associated with the wave operator under outgoing boundary conditions, as described by Dyatlov and Zworski \cite{Dyatlov_2011, dyatlov2019mathematical}. They can also be cast as the (proper) eigenvalues of a non-selfadjoint operator defined in the appropriate Hilbert space \cite{Warnick:2013hba,Ansorg_2016}. 
The two characterisations are not exactly equivalent, since quasi-normal modes defined as eigenvalues
do include scattering resonances but potentially also other `resonant' frequencies 
 (cf. section 7 in \cite{Warnick:2013hba} for a discussion).
This eigenvalue
characterisation is more akin to standard treatment in optics, where quasi-normal modes  heuristically identified with operator eigenvalues.

In general, QNMs are associated with complex frequencies: the real part encodes oscillatory behavior linked to energy confined within the cavity, while the imaginary part reflects temporal attenuation due to leakage. Despite extensive study, many questions remain unresolved—most notably, the completeness of QNMs. Unlike self-adjoint operators, which admit a spectral theorem and yield to normal modes. Non-selfadjoint operators lack a comparable theory. 
In this work, we address the problem of determining QNMs in a one-dimensional optical setting by integrating complementary analytical and computational approaches, with a focus on the analysis of spectral 
QNM stability in optics (as an example of such phenomena, cf \cite{wu2025reflections} for a study of the spectral instability under far-disctance perturbations). Although this combination may initially appear intricate, it profits from key insight into the spectral properties of non-selfadjoint operators. A central advantage of our framework is the natural non-divergence of modes within hyperboloidal compactified coordinates, therefore normalization is not a problem. Using this approach, we identify a family of purely imaginary eigenfrequencies. As commented above, eigenvalues can contain additional frequencies to scattering resonances ones, and the
specific assessment of such purely imaginary frequencies will be the subject of future work.

Our method relies on three principal components, each essential for solving the eigenvalue problem under outgoing boundary conditions. 
\begin{enumerate}[ i.]
\item A compactification along hyperboloidal slices through a coordinate transformation motivated by gravitational-wave studies in black hole spacetimes, incorporating outgoing boundary conditions in a geometrically natural manner, as regularity conditions at the boundaries.
\item Adopting a  Lorentz model for permittivity and introducing two auxiliary fields, a well-established technique here implemented within hyperboloidal slicing coordinates. 
\item Implementing numerical spectral methods based on Chebyshev polynomials to compute the resulting eigenvalue problem with high accuracy.
\end{enumerate}
In addition, we discuss the computation of the $\epsilon-$pseudospectrum to analyze the stability of the obtained QNMs, providing a systematic framework for assessing resonance instability. This is akin to the strategy in
\cite{ColRomHan19} for  the estimation of errors
in the numerical calculation of spectra.

The article is organized as follows. Section~II introduces the mathematical formulation and tools underlying our approach. Section~III discusses the numerical results of the spectrum calculations. Section~IV develops the theoretical framework of pseudospectrum analysis and explains its relevance to non-selfadjoint resonance problems. with illustrative numerical examples, demonstrating both the spectral method and the pseudospectrum. Finally, Section~V summarizes our conclusions and outlines directions for future research, with an emphasis on the broader implications for photonic and electromagnetic systems.
Appendices are dedicated to give more details on Drude model, all Chebyshev polynomials-related equations used in the numerical translation of the problem, and to hyperboloidal slices approach.

\section{Approaches and tools}
This section presents the mathematical formulation and numerical tools used in our approach. We consider a one-dimensional open optical cavity modeled as a compactly supported scatterer. The wave equation that describes light scattering in the cavity:

\begin{equation}
\label{gen_wave_eq}
[\epsilon(\omega, x) \omega^2+ \partial_x^2 ] \phi(\omega, x) = S(\omega, x),
\end{equation}
where $S(\omega, x)$ is a source term that should be set to $0$ in order to find the modes of resonance.
$\epsilon$ is the permittivity.

In order to describe the problem as an evolution problem in space-time, it is necessary to write the equations with the fields as functions of time and space. This is by taking the inverse Fourier transformation in time of the source-free Eq. \ref{gen_wave_eq}
\begin{equation}
\label{gen_wave_eq_t}
[- \epsilon(t) *  \partial_t^2 + \partial_x^2] \phi(t, x) = 0,
\end{equation}
where  $*$ denotes convolution. Handling this convolution in the hyperboloidal coordinates needs special attention \cite{10.3389/fphy.2024.1457543}, instead, we adopt a dispersive permittivity model of Lorentz–Drude type. This model captures the dispersive and absorptive behavior of optical materials and is widely used for metals and dielectrics. We assume that the cavity and its exterior are both described by this model, with the possibility that the exterior medium may have different material or parameters on each side of the cavity.
Thus the frequency–dependent permittivity in the cavity or in each of the exterior regions takes the form
\begin{equation}
\label{Permi_f}
\varepsilon(\omega) = \varepsilon_\infty - \frac{\omega_p^2}{\omega^2 + i \Gamma \omega - \omega_0^2},
\end{equation}
where $\varepsilon_\infty$ denotes the high-frequency permittivity, $\omega_p$ the plasma frequency, $\Gamma$ the damping parameter, and $\omega_0$ the resonance frequency.

The physical mechanism behind this model in space-time is described by Lorentz model. Using the space-time version with auxiliary field formulation replaces the convolution with a system of diferential equations, yielding a form well suited for the hyperboloidal transformation and subsequent discretization. 
In this model, the electron–nucleus interaction is represented by a damped-forced harmonic oscillator. The equation governing the displacement $r$ of the electron from the nucleus is
\begin{equation}
\label{LorentzM}
\frac{\partial^2 \mathbf{r}}{\partial t^2}+ \Gamma \frac{\partial \mathbf{r}}{\partial t}+\omega_0^2 \mathbf{r}=\frac{e\mathbf{E}}{m},
\end{equation}
where $\Gamma$ is the damping rate associated with absorption, $\omega_0$ is the characteristic frequency of the oscillator, $m$ is the electron mass, and $\mathbf{E}$ is the external electric field.  

Defining the polarization as $\mathbf{P}=Ne\mathbf{r}$, the equation can be rewritten in terms of $\mathbf{P}$ as
\begin{equation}
\label{polarization eq}
\frac{\partial^2 \mathbf{P}}{\partial t^2}+ \Gamma \frac{\partial \mathbf{P}}{\partial t}+\omega_0^2 \mathbf{P}=\omega_p^2 \epsilon_{\infty} \mathbf{E},
\end{equation}
where $\omega_p^2=\frac{Ne^2}{m \epsilon_{\infty}}$ and $N$ is the electron density. From Maxwell’s equations, we obtain
\begin{equation}
\label{maxwell}
\begin{array}{lllllllll}
\displaystyle
\frac{\partial \mathbf{H}}{\partial t}=- \frac{1}{\mu_0} \nabla \times \mathbf{E}, \\
\displaystyle
\frac{\partial \mathbf{E}}{\partial t}= \frac{1}{\epsilon_{\infty}} \bigg(\nabla \times \mathbf{H} -\frac{\partial \mathbf{P}}{\partial t}\bigg),
\end{array}
\end{equation}
which leads to the following equation for $\mathbf{E}$:

\begin{equation}
\label{Helmoheltz}
\frac{\partial^2 \mathbf{E}}{\partial t^2}= \frac{1}{\epsilon_{\infty}} \left(- \frac{1}{\mu_0} \bigtriangleup \mathbf{E} - \frac{\partial^2 \mathbf{P}}{\partial t^2}\right).
\end{equation}

We focus here on one-dimensional system with $\mathbf{P}$ is aligned with $\mathbf{E}$, so we can write Eqs.~\eqref{polarization eq} and \eqref{Helmoheltz} in scalar form. $E$ and $P$ are denoted to the scalar versions of these two fields.

 \begin{gather}
\label{Eq_grp}
\frac{\partial^2 E}{\partial t^2}= \frac{1}{\epsilon_{\infty}} \left(- \frac{1}{\mu_0}  \frac{\partial^2 E}{\partial x^2} - \frac{\partial^2 P}{\partial t^2}\right), \\
\label{Eq_grp2}
\frac{\partial^2 P}{\partial t^2}+ \Gamma \frac{\partial P}{\partial t}+\omega_0^2 P=\omega_p^2 \epsilon_{\infty} E.
\end{gather}
To solve these equations in different regions of space (inside and outside the cavity), the next step will be to transform them to equations in a hyperboloidal scheme. 
Next subsection introduces the hyperboloidal approach which will help not only in compactifying space, but also to formulate the electromagnetic problem in a Hilbert space. 

\subsection{Hyperboloidal slices approach}
A major difficulty in QNM analysis is the divergence of the modal fields at spatial infinity. Several strategies have been developed to address this issue, largely motivated by computational and practical considerations. For instance, perfectly matched layers (PMLs) are widely used in numerical simulations \cite{LalYanVyn17}, while alternative techniques based on complex scaling have also been explored  (see  \cite{zenginouglu2026null,wess2026finite}
for a discussion of the relation of these methods with the hyperboloidal slicing scheme).

\begin{figure}[h!]
\centering
\includegraphics[width=8.5cm]{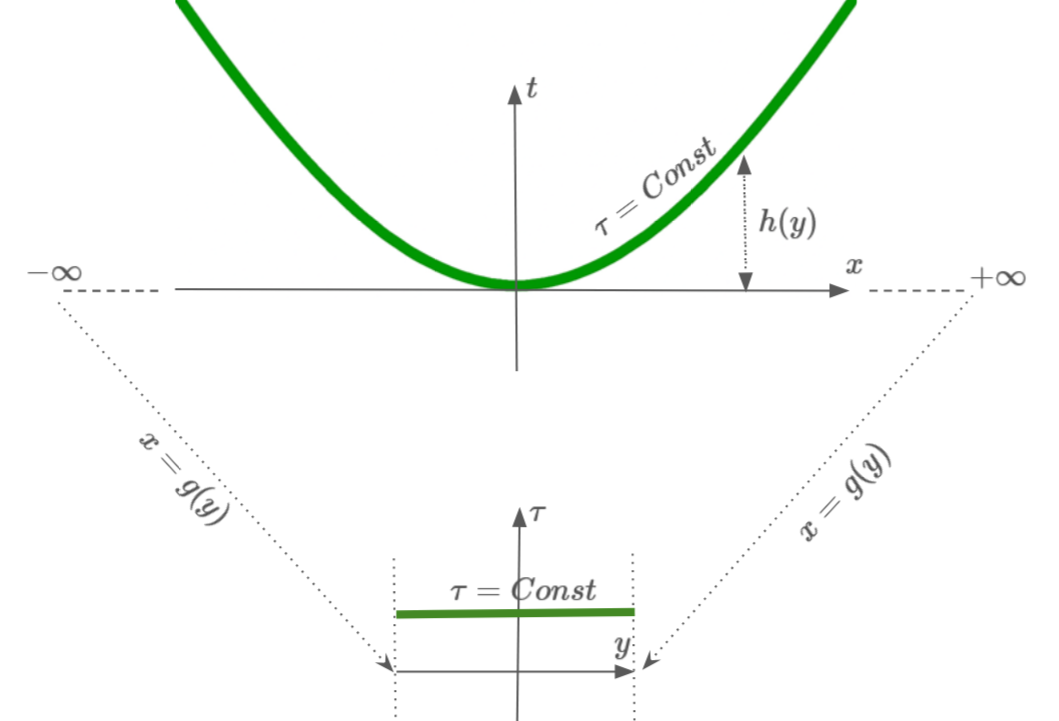}
\caption{Schematic representation of the hyperboloidal coordinate transformation in Eq.~\eqref{transformation_org}. 
\emph{Top panel:} $(t,x)$ coordinates, where the height function $h(y)$ bends time slices so that future null infinity is reached as $x \rightarrow \pm \infty$. 
\emph{Bottom panel:} Hyperboloidal coordinates $(\tau,y)$, where $g(y)$ maps the infinite domain $x \in (-\infty,\infty)$ onto the finite interval $y \in (a,b)$. 
The added boundary points $a$ and $b$ represent future null infinity $\mathscr{I}^+$.}
\label{fig:HypCoordTransf}
\end{figure}

In this work we adopt the \emph{hyperboloidal slicing} approach, which tackles the divergence problem at its root by introducing a time coordinate whose level sets are asymptotically outgoing. The central idea is geometric: rather than truncating the domain and approximating radiation conditions, we choose coordinates in which the outer boundary of the numerical domain coincides with an asymptotic outgoing boundary. Figure~\ref{fig:HypCoordTransf} provides an intuition for what is meant by \emph{null infinity}---the asymptotic endpoint of outgoing characteristics. Our implementation rely on adopting hyperboloidal time slices and \emph{compactify only the spatial coordinate along each slice}, thereby mapping the space to a finite interval.

Concretely, we introduce new coordinates $(\tau,y)$ of the form \cite{Ansorg_2016,Macedo_2018,Macedo_2020,Zengino_lu_2011}
\begin{equation}
\label{transformation_org}
\begin{array}{llll}
t = \tau - h(y),\\
x= g(y),
\end{array}
\end{equation}
where $h$ is a \emph{height function} that bends the time slices and $g$ is a \emph{compactification map} that sends the infinite physical domain to a bounded interval in $y$. With an appropriate choice of $(h,g)$, the hypersurfaces $\tau=\mathrm{const}$ become asymptotically characteristic and intersect future null infinity, so that the outgoing boundary condition is enforced at the compactified boundaries. A schematic representation of this construction adapted from \cite{Jaramillo:2020tuu} is shown in Fig.~\ref{fig:HypCoordTransf}.

A crucial point in electromagnetic and dispersive settings is that, unlike relativistic wave equations, there is no unique preferred propagation speed. As emphasized in the discussion of Burgess \& K{\"o}nig \cite{10.3389/fphy.2024.1457543}, in dispersive systems the relevant outgoing direction is controlled by the \emph{asymptotic group velocity}, and no single coordinate transformation can make all possible outgoing behaviors simultaneously bounded. In practice, one therefore parametrizes the asymptotics of $h(y)$ so that $\tau=\mathrm{const}$ surfaces align, at large $|x|$, with outward trajectories associated with a chosen asymptotic group velocity. This choice enforces the desired boundary behavior for the corresponding class of modes at a given frequency, although it does not by itself guarantee that such modes exist.

In the present work we deal with this problem by restricting attention to \emph{compactly supported resonators}, i.e., cavities described by material inhomogeneities confined to a finite region and embedded in a homogeneous exterior. In contrast with dispersive situations in modified theories of gravity, on which Burgess and Kronig focus on, such a tratament is natural in a large class of optical resonators. This assumption cleanly separates the interior dispersive structure from a uniform asymptotic medium, making the hyperboloidal construction both physically meaningful and numerically robust. It also motivates (i) the specific asymptotic form adopted for the height function $h(y)$, which is tuned to the exterior outgoing propagation, and (ii) the use of multi-grids: different types of spectral grids: Lobatto in the middle covering the cavity to include its both endpoints,  Radau-type spectral grids (left and right Radau grids): to include the endpoints of the cavity and apply continiuity conditions while not including the infinity but dealing with it as the asymptotic outgoing endpoints, this way the outer boundary conditions are imposed and resolved sharply within a Chebyshev discretization. Appendix \ref{ChebyApp} contains more details about different grids used in Chebyshev spectral methods.

With a suitable choice of $h(y)$ and $g(y)$, one recovers the Bizo\'n--Mach coordinates \cite{Bizon:2014nla,Donninger:2020sqm}:
\begin{equation}
\label{transformation}
\begin{array}{llll}
\tau= t-\ln (2 \cosh(x)),\\
y=\tanh(x).
\end{array}
\end{equation}
The second line compactifies the spatial domain, since $y \in ]-1,1[$ corresponds to $x \in ]-\infty,\infty[$. Moreover, as $x \to +\infty$ we recover $\tau=t-x$, while as $x \to -\infty$ we obtain $\tau=t+x$, i.e., hyperboloidal slices that asymptotically approach outgoing characteristics.

From Eq.~\eqref{transformation} we obtain the following relations:
\begin{equation}
\label{tangent_vectors}
\begin{array}{ccc}
\begin{aligned}
\partial_{t} = & \partial_{\tau}, \\
\partial_{x} = & -y \partial_{\tau} + (1-y^2) \partial_{y}, \\
\partial_{t}^2 = & \partial_{\tau}^2, \\
\partial_{x}^2 = & y^2 \partial_{\tau}^2  -2y (1-y^2) \partial_{\tau} \partial_{y} - (1-y^2) \partial_{\tau} \\
& \ - 2y (1-y^2) \partial_y + (1-y^2) \partial_y^2.
\end{aligned}
\end{array}
\end{equation}

\noindent
Note that the first equation in \eqref{tangent_vectors} indicating the stationarity of the system. Having established the hyperboloidal compactification framework, we now embed the physical system into it. Specifically, we describe electromagnetic wave propagation in dispersive media using the Maxwell--Lorentz model, reformulated in hyperboloidal coordinates so that outgoing boundary conditions are incorporated at the compactified boundaries.

\subsection{One-dimensional electromagnetic problem in hyperboloidal slices}

To incorporate the hyperboloidal slicing approach into electromagnetism,
we write Eq.~\eqref{Eq_grp} and Eq.~\eqref{Eq_grp2} in hyperboloidal coordinates by using Eq.~\eqref{tangent_vectors}. This gives

\begin{equation}
	\label{EM in HP1}
	\begin{aligned}
		& \partial_{\tau}^2  P  + \Gamma \partial_{\tau} P + \omega_0^2 P = \epsilon_{\infty} \omega_p^2 E, \\
		& (\epsilon_{\infty}  \mu_0  - y^2) \partial_{\tau}^2 E + 2y (1-y^2) \partial_y \partial_{\tau} E \\
		& \quad + (1-y^2) \partial_{\tau} E  +2y (1-y^2) \partial_y E  - (1-y^2)^2 \partial_y^2 E \\
		& \quad + \mu_0  \partial_{\tau}^2 P  = 0.
	\end{aligned}
\end{equation}

To simplify this system, we perform a first-order reduction in time by introducing the auxiliary scalar fields $A$ and $B$, defined by

\begin{equation}
	\label{auxiliary A B}
	\begin{array}{lllllll}
		\partial_{\tau} P=A, \\
		\partial_{\tau} E=B,
	\end{array}
\end{equation}

Substituting these definitions into Eq.~\eqref{EM in HP1} gives

\begin{equation}
	\label{EM in HP}
	\begin{aligned}
		& \partial_{\tau} A  + \Gamma A + \omega_0^2 P = \epsilon_{\infty} \omega_p^2 E, \\
		& (\epsilon_{\infty}  \mu_0  - y^2) \partial_{\tau} B + 2y (1-y^2) \partial_y B \\
		& \quad + (1-y^2) B +2y (1-y^2) \partial_y E  - (1-y^2)^2 \partial_y^2 E \\
		& \quad + \mu_0  \partial_{\tau} A  = 0.
	\end{aligned}
\end{equation}

We then apply the Fourier transformation. Since the hyperboloidal coordinate transformation satisfies $\partial_t=\partial_\tau$, the physical and hyperboloidal frequencies coincide, i.e.,
\[
i\omega_t=i\omega_\tau.
\]
Equation~\eqref{EM in HP} therefore becomes

\begin{equation}
	\label{EM in HP Fourrier}
	\begin{aligned}
		& i \omega P = A, \\
		& i \omega E = B, \\
		& i \omega A = - \Gamma A -\omega_0^2 P + \omega_p^2 \epsilon_{\infty} E, \\
		& i \omega (\epsilon_{\infty}  \mu_0  - y^2) B + i \omega \mu_0 A = - 2y (1-y^2) \partial_y B \\
		& \quad - (1-y^2) B - 2y (1-y^2) \partial_y E  + (1-y^2)^2 \partial_y^2 E.
	\end{aligned}
\end{equation}

Throughout this work we adopt the Fourier convention
\[
\partial_t \mapsto i\omega,
\]
so that time derivatives translate into multiplication by $i\omega$. In this convention, the eigenvalues $\lambda$ of the operator and the physical eigenfrequencies $\omega$ are related by
\[
\lambda=i\omega.
\]

The spatial differential operators resulting from rewriting the Maxwell--Lorentz equations in hyperboloidal coordinates are

\begin{equation}
	\begin{aligned}
		L_1 &= -2y(1-y^2)\partial_y +(1-y^2)^2\partial_y^2,\\
		L_2 &= -2y(1-y^2)\partial_y -(1-y^2)I.
	\end{aligned}
\end{equation}

Equation~\eqref{EM in HP Fourrier} can then be written in the generalized matrix form

\begin{equation}
	\label{matricial form EM}
	\begin{aligned}
		& \begin{bmatrix}
			L_1 & L_2 &0 &0 \\
			0& 1 &0 & 0 \\
			0& 0 & 0 & 1 \\
			\omega_p^2 \epsilon_{\infty} & 0 &-\omega_0^2 & -\Gamma
		\end{bmatrix}
		\begin{bmatrix}
			E\\
			B\\
			P\\
			A
		\end{bmatrix}  \\
		= i \omega
		& \begin{bmatrix}
			0 & \epsilon_{\infty} \mu_0 I - y^2 & 0 & \mu_0 \\
			1 &0 & 0 &0\\
			0& 0 &  1 & 0\\
			0 &0 & 0 &1
		\end{bmatrix}
		\begin{bmatrix}
			E\\
			B\\
			P\\
			A
		\end{bmatrix},
	\end{aligned}
\end{equation}

Equation~\eqref{matricial form EM} defines the continuous generalized eigenvalue problem in the hyperboloidal coordinate.

To solve this problem numerically, we first decompose the physical domain into three regions,
\begin{equation}
	]-\infty,a], \qquad [a,b], \qquad [b,+\infty[,
\end{equation}
where the central region contains the optical cavity while the two exterior regions represent the surrounding homogeneous medium. The hyperboloidal transformation
\begin{equation}
	y=\tanh x
\end{equation}
compactifies these three physical regions into
\begin{equation}
	]-1,\tanh a], \qquad
	[\tanh a,\tanh b], \qquad
	[\tanh b,1[,
\end{equation}
respectively. The endpoints $y=-1$ and $y=1$ therefore correspond to the two asymptotic infinities, whereas $y=\tanh a$ and $y=\tanh b$ correspond to the physical interfaces of the cavity.

Since the spectral discretization is constructed on the standard Chebyshev interval $[-1,1]$, each of these three hyperboloidal subdomains is subsequently mapped independently onto a Chebyshev reference coordinate $\xi_j\in[-1,1]$. We thus employ a multi-domain spectral discretization consisting of three Chebyshev grids, one for each physical region.

Importantly, no explicit outgoing boundary conditions are required at $y=\pm1$, since the exterior hyperboloidal slices extend to null infinity and therefore encode the correct radiative behavior~\cite{Jaramillo:2020tuu}. The treatment of these asymptotic endpoints determines the choice of the Chebyshev grids, as described below.
\vspace{-0.7cm}

\subsection{Numerical Chebyshev spectral methods}

We now turn to the numerical implementation of the eigenvalue problem derived above, using spectral methods as the principal computational tool. These methods provide the accuracy and efficiency required for solving non-selfadjoint problems with outgoing boundary conditions.  
For non-periodic systems, algebraic polynomial bases such as Chebyshev polynomials are required.
Our numerical implementation is essentially inspired and following the detailed work of Marcus Ansorg (2013) \cite{Marcus} about spectral methods, also we refer to the comprehensive treatment in Canuto, Hussaini, Quarteroni, and Zang \cite{canuto2007spectral} (authors of an earlier landmark book in 1988), and the widely used text by Trefethen \cite{trefethen2000spectral}, which introduced practical implementations and software.

\begin{figure}[H]
	\centering
	\includegraphics[scale=0.21]{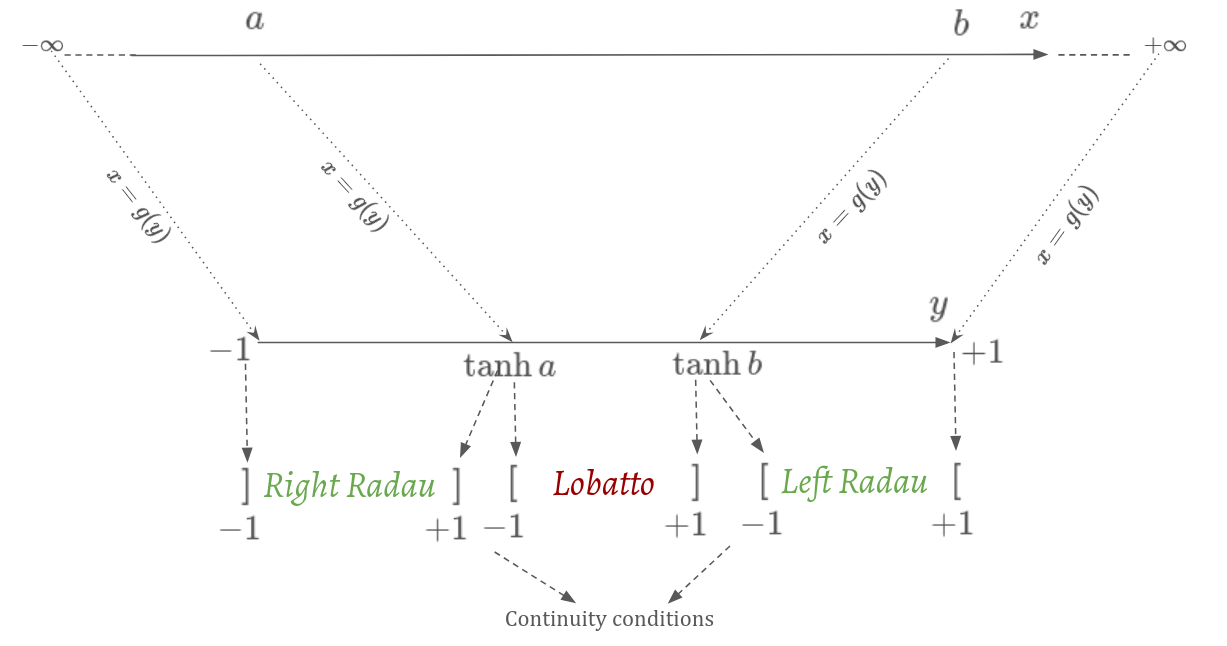}
	\caption{Mapping from $x$ coordinate to hyperboloidal space coordinate $y$ and then mapping each of the three regions to a different Chebyshev grid }
	\label{multigrids}
\end{figure}

We discretize the generalized eigenvalue problem using a multi-grid Chebyshev collocation method. The three hyperboloidal subdomains
\begin{equation}
	\begin{aligned}
		y_1 &\in ]-1,\tanh a],\
		y_2 &\in [\tanh a,\tanh b],\
		y_3 &\in [\tanh b,1[,
	\end{aligned}
\end{equation}
are each mapped onto the standard Chebyshev interval $[-1,1]$. That is illustrated in Fig. \ref{multigrids}.

The central region has two finite physical interfaces, at $y=\tanh a$ and $y=\tanh b$. Both endpoints must therefore be included in the discretization, and a Chebyshev--Lobatto grid is used in this region. In contrast, each exterior region contains one finite interface and one endpoint corresponding to null infinity. We therefore use Chebyshev--Radau grids, which include the finite interface while excluding the asymptotic endpoint. More precisely, we use:
\begin{itemize}
	\item A right-Radau grid in the left exterior region, including $y=\tanh a$ and excluding $y=-1$;
	\item A Lobatto grid in the cavity region, including both $y=\tanh a$ and $y=\tanh b$;
	\item A left-Radau grid in the right exterior region, including $y=\tanh b$ and excluding $y=1$.
\end{itemize}

This choice gives three independent spectral grids while retaining the interface points required to couple adjacent regions. The generalized eigenvalue problem~\eqref{matricial form EM} is then written and discretized independently on each grid using the material parameters appropriate to the corresponding region.

If each grid contains $N+1$ collocation points, the state vector in each subdomain is written as

\begin{equation}
	\psi_{y_j}=
	\begin{bmatrix}
	E_{y_j}\\
	B_{y_j}\\
	P_{y_j}\\
	A_{y_j}	
	\end{bmatrix},
	\qquad j=1,2,3,
\end{equation}
where $E_i$, $B_i$, $P_i$, and $A_i$ denote the vectors of
field values evaluated at the collocation points of the $i$th
subdomain. Hence, $\psi_i$ has dimension $4(N+1)$. The three regional generalized eigenvalue problems can then be written as
\begin{equation}
		\label{three domaines}
		\begin{aligned}
			L_{y_1}\psi_{y_1} &= i\omega M_{y_1}\psi_{y_1},\\
			L_{y_2}\psi_{y_2} &= i\omega M_{y_2}\psi_{y_2},\\
			L_{y_3}\psi_{y_3} &= i\omega M_{y_3}\psi_{y_3}.
		\end{aligned}
	\end{equation}
where the difference between the three problems lies in the corresponding material parameters and spectral grids.

Before assembling the three regional problems into a single global system, continuity conditions are imposed at the two physical interfaces. Denoting
\begin{equation}
	y_a = \tanh a,
	\qquad
	y_b = \tanh b,
\end{equation}
the electric field and its derivative with respect to the hyperboloidal coordinate $y$ satisfy
\begin{equation}
	\label{continuity}
	\begin{gathered}
		E_{y_1}(y_a) = E_{y_2}(y_a),
		\\[1.5ex]
		E_{y_2}(y_b) = E_{y_3}(y_b),
		\\[1.5ex]
		\left.
		\frac{\mathrm{d}E_{y_1}}{\mathrm{d}y}
		\right|_{y=y_a}
		=
		\left.
		\frac{\mathrm{d}E_{y_2}}{\mathrm{d}y}
		\right|_{y=y_a},
		\\[1.5ex]
		\left.
		\frac{\mathrm{d}E_{y_2}}{\mathrm{d}y}
		\right|_{y=y_b}
		=
		\left.
		\frac{\mathrm{d}E_{y_3}}{\mathrm{d}y}
		\right|_{y=y_b}.
	\end{gathered}
\end{equation}
After mapping each subdomain independently onto its local Chebyshev coordinate $\xi_j\in[-1,1]$, the derivative conditions are implemented using the corresponding rescaled differentiation matrices. In each subdomain,
\begin{equation}
	\frac{\mathrm{d}}{\mathrm{d}y}=
	\frac{\mathrm{d}\xi_j}{\mathrm{d}y}
	\frac{\mathrm{d}}{\mathrm{d}\xi_j},
\end{equation}
so that the different affine rescalings of the three subdomains are taken into account when matching the derivatives across the interfaces.
In practice, these continuity conditions are enforced by replacing the corresponding rows of the discretized regional matrices. We denote the resulting matrices, after imposing the interface conditions, by $L'_{y_1}$, $L'_{y_2}$, $L'_{y_3}$ and $M'_{y_1}$, $M'_{y_2}$, $M'_{y_3}$.

The three modified regional problems are then assembled into the global generalized eigenvalue problem,
\begin{equation}
	\label{ev_prob_op}
	\begin{split}
		&
		\begin{bmatrix}
			L'_{y_1} & 0 & 0 \\
			0 & L'_{y_2} & 0 \\
			0 & 0 & L'_{y_3}
		\end{bmatrix}
		\begin{bmatrix}
			\psi_{y_1} \\
			\psi_{y_2} \\
			\psi_{y_3}
		\end{bmatrix}
		\\[1ex]
		&\qquad =
		i\omega
		\begin{bmatrix}
			M'_{y_1} & 0 & 0 \\
			0 & M'_{y_2} & 0 \\
			0 & 0 & M'_{y_3}
		\end{bmatrix}
		\begin{bmatrix}
			\psi_{y_1} \\
			\psi_{y_2} \\
			\psi_{y_3}
		\end{bmatrix}.
	\end{split}
\end{equation}
The dimensions of the resulting system is a
$\big(3 \times 4\times (N+1)\big)\times
\big(3 \times 4\times (N+1)\big)$ matrix
and defines the global generalized eigenvalue problem used for the numerical computation.
\section{Spectrum results}
\subsection{QNM eigenfrequencies and Convergence results}
\begin{figure}[H]
\centering
\includegraphics[scale=0.7]{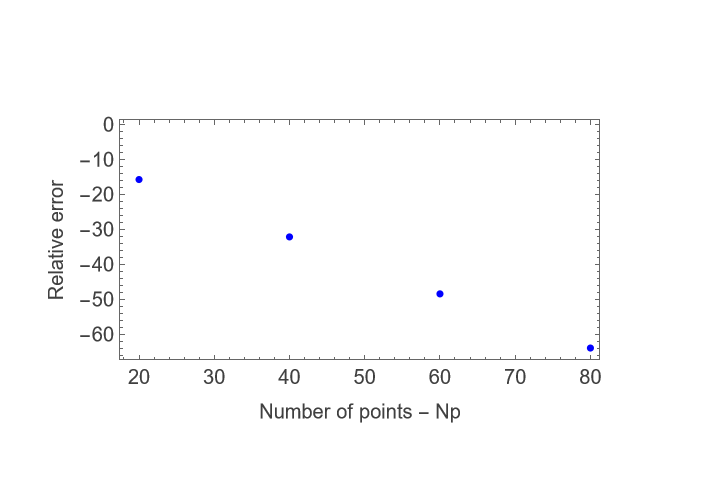}
\caption{Convergence of quasi-normal mode eigenfrequencies as a function of number of points considered for the descretization.
The results are for a particular eigenvalue ~$ 2.66849 + 1.754697 i$ confirm the accuracy of the Chebyshev discretization scheme, with stabilization of the computed eigenvalues as the number of collocation points increases.}
\label{conv}
\end{figure}

Figure~\ref{EV} illustrates the spectrum obtained numerically from Eq.~\eqref{ev_prob_op} with continuity conditions imposed via Eq.~\eqref{continuity}. Results are shown for the case $\epsilon_{\infty}=1$, $\omega_p=1$, $\Gamma=0.1$, and $\omega_0=0$.
To show the convergence of eigenvalues, we focus on one of the calculated eigenvalues.
We choose an eigenvalue approximated by $ 2.66849 + 1.754697 i$, track it for repeated calculations for increased number of collocation points, results in Fig.~\ref{conv}.
We define a relative error  $\delta_n^{N}$ as in Eq.~\eqref{eq:conv_error}.

\begin{equation}
\label{eq:conv_error}
\delta_n^N = \log \left| 1 - \frac{\omega_{n}}{\omega{N}} \right| ,
\end{equation}

where $\omega_n$ is the frequency approximated by ($ 2.66849 + 1.754697 i$) for grids with $n+1$ point. $\omega_N$ is for a grid with $N$ point, where $N$ here represents the maximum number of points used for these calculations.
Figure~\ref{conv} presents the logarithmic relative error $\delta_n^{(N)}$ as a function of the number of grid points ($n$). 
A rapid convergence is observed, even for relatively small number of points.

The eigenfrequencies used in Fig.~\ref{conv} are obtained from 
Eq.~\eqref{ev_prob_op} together with the continuity conditions in 
Eq.~\eqref{continuity}.  

\subsection{Spectrum and "antibound" states}
\begin{figure}[H]
\centering
\includegraphics[scale=0.9]{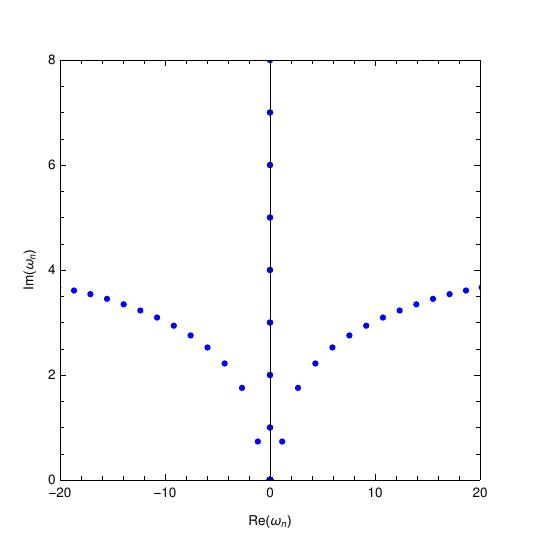}
\caption{Spectrum of the generalized eigenvalue problem \eqref{ev_prob_op} with continuity conditions \eqref{continuity}. 
Parameters are $\epsilon_{\infty}=1$, $\omega_p=1$, $\Gamma=0.1$, and $\omega_0=0$. 
Two distinct families of eigenvalues are observed: (i) a discrete set lying purely on the imaginary axis, $\omega_m = i m$, $m \in \mathbb{N}$, corresponding to so-called ``antibound'' states; and (ii) a logarithmically organized branch of complex frequencies, representing the physical quasi-normal modes (QNMs). 
This clear separation highlights the coexistence of non-oscillatory purely imaginary modes with oscillatory decaying QNMs in the same open optical system.}
\label{EV}
\end{figure}

Two distinct families of eigenvalues in Fig.~\ref{EV} are identified:
\begin{itemize}
\item[i)] Eigenvalues distributed along the imaginary axis, $\omega_m = i m$, with $m \in \mathbb{N}$.  
\item[ii)] Eigenvalues following a logarithmic curve, which correspond to the QNMs.  
\end{itemize}

\subsubsection{Antibound states: Purely imaginary eigenfrequencies}
We now examine the purely imaginary eigenfrequencies and verify that they correspond to genuine solutions of the eigenvalue problem rather than numerical artifacts.  

Outgoing boundary conditions are designed to enforce the Sommerfeld radiation condition at infinity, ensuring that no waves enter the system from infinity. Consequently, in regions I and III (free-space regions), no incoming wave is allowed, since free space does not generate incoming waves in the absence of permittivity variations or external potentials.  

For example, in region III the general solution takes the form
\[
\phi_{III}=A_3 e^{-i\omega x}+B_3 e^{i \omega x}.
\]
Here, the term proportional to $A_3$ represents the outgoing wave, while the term proportional to $B_3$ represents the incoming wave. To satisfy outgoing boundary conditions, $B_3$ must vanish, except in the special case where $e^{i \omega x}$ does not correspond to a propagating mode. This occurs precisely when the real part of $\omega$ vanishes. In such cases, with $\omega^I>0$, this term decays as $x\to+\infty$.  

Thus, a solution in region III of the form
\[
\phi_{III}=A_3 e^{m x}+B_3 e^{-m x}, \qquad \omega^I = m,
\]
remains consistent with outgoing boundary conditions and satisfies the free-space wave equation.  

Let us now examine the same solution in hyperboloidal coordinates. A field of the form $\phi(t,x) = e^{i \omega t} \phi(x)$ becomes
\[
\phi(\tau,y) = e^{i \omega \tau} \phi_y(y),
\]
where the transformation relations give
\[
\phi_y(y)= \phi(x) \, e^{i \omega \ln(2 \cosh(x))}.
\]
Since $y = \tanh(x)$, we have $x=\ln \sqrt{\tfrac{1+y}{1-y}}$ and $2 \cosh(x) = \tfrac{2}{\sqrt{1-y^2}}$.  

The general free-space solution
\[
\phi(x) = A\, e^{-i\omega x} + B\, e^{i \omega x}
\]
thus yields
\begin{align}
\phi_y(y)
&= \big( A\, e^{-i\omega x} + B\, e^{i\omega x} \big)\,
     e^{i\omega \ln(2\cosh x)} \notag \\
&= A\, f_-(y) + B\, f_+(y),
\end{align}
where
\[
f_\pm(y) =
    \left(\sqrt{\tfrac{1+y}{1-y}}\right)^{\pm i\omega}
    \left( \tfrac{2}{\sqrt{1-y^2}} \right)^{i\omega}
    = 2^{i\omega}(1\pm y)^{-i\omega}.
\]
Thus,
\[
\phi_y(y) = 2^{i\omega} \big[ A\,(1+y)^{-i\omega} + B\,(1-y)^{-i\omega} \big].
\]

In order for the last expression to define an analytic solution, the imaginary part of $\omega$ must be positive and integer. Hence purely imaginary frequencies $\omega = i m$, $m \in \mathbb{N}$, naturally arise.  
The physical interpretation of these eigenvalues remains subtle. Similar eigenvalues have been reported in several earlier works~\cite{HokNob65,OhaGin74,Belchev2011,Siegert39} and are often referred to as \emph{antibound} (or \emph{anti-bounded}) states. Gamow~\cite{Gamow28} already suggested that, although such frequencies may not correspond to physical modes, their eigenfunctions can nonetheless contribute to resonant expansions~\cite{Siegert39,Peierls59,Belchev2011}, thereby potentially influencing the temporal decay of fields.  

\noindent
In the current optical setting, the precise role of these purely imaginary frequencies remains unresolved. It is not yet clear whether these purely imaginary eigenvalues contribute meaningfully to the scattered-field expansion or if they are best regarded as model artifacts. Future investigations, particularly in the time domain, should test their role as additional decay contributors, and if they are necessary to the expansion or not. Such studies will clarify whether purely imaginary modes constitute a fundamental resonance feature of open optical systems, or whether they can be safely neglected in practice.

\subsubsection{Regge QNM branches: Logarithmic asymptotics}

The quasi-normal mode spectrum exhibits a distinctive asymptotic structure:
\begin{itemize}
\item The imaginary parts of the eigenfrequencies scale logarithmically with their real parts.  
\item The spacing between consecutive real parts approaches a constant in the high-frequency limit.  
\end{itemize}

This organization of the spectrum, commonly referred to as the \textit{Regge branches} of QNMs, is closely connected to Weyl-type asymptotics. In particular, the convergence of the real-part spacing reflects the same spectral regularity that underlies Weyl’s law for eigenvalue counting, here generalized to a non-selfadjoint open-system setting.  

\begin{figure}[H]
\centering
\includegraphics[scale=0.7]{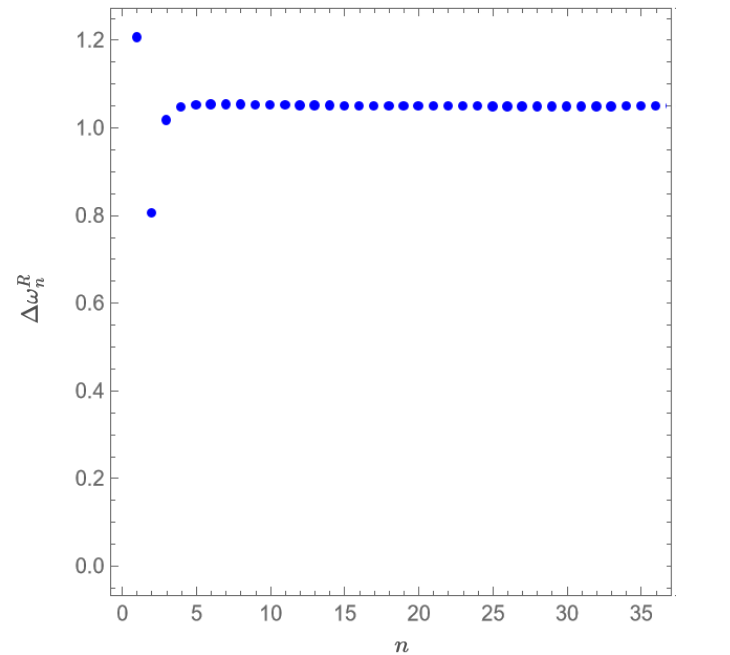}
\caption{Asymptotic spacing of QNM eigenfrequencies. The plot shows the difference between successive real parts $\Delta \omega^R_n = \omega^R_{n+1} - \omega^R_n$ as a function of mode index $n$, with $\omega^R_n$ ordered increasingly. The rapid convergence toward a constant demonstrates the logarithmic Regge-type structure, consistent with Weyl-type asymptotics for resonance distributions.}
\label{Diff_r}
\end{figure}

\subsection{QNM eigenfunctions: Normalization}
\begin{figure}[H]
\centering
\includegraphics[scale=0.6]{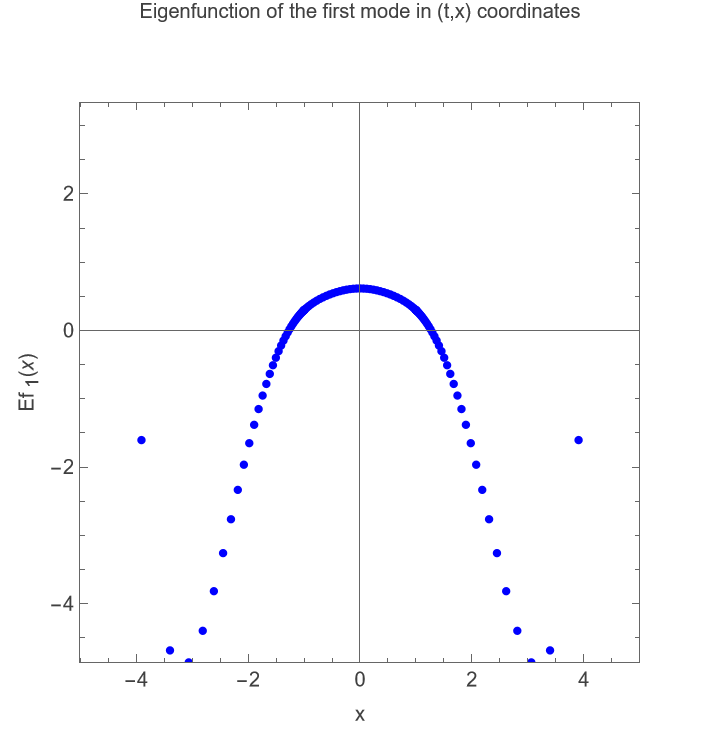}
\caption{Fundamental QNM eigenfunction plotted as a function of the physical coordinate $x$. 
The field diverges as $|x|\to\infty$, reflecting the well-known non-normalizability of QNMs in standard spacetime coordinates. 
Two conjugate eigenfunctions are shown, corresponding to $\omega_f$ and $- \omega_f^*$, which are identical in magnitude but differ by complex conjugation.}
\label{Eftx}
\end{figure}

\begin{figure}[H]
\centering
\includegraphics[scale=0.6]{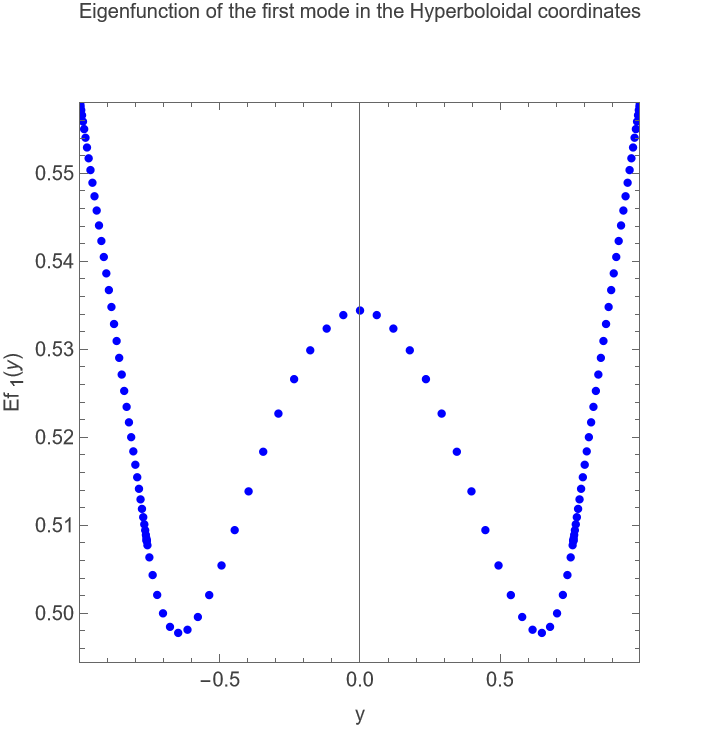}
\caption{Fundamental QNM eigenfunction plotted in compactified hyperboloidal coordinates $y=\tanh(x)$. 
In this representation the mode is square-integrable, allowing rigorous normalization. 
As in Fig.~\ref{Eftx}, both eigenfunctions associated with $\omega_f$ and $- \omega_f^*$ are shown. 
The compactification thus enables the consistent computation of mode volumes and other physically relevant quantities.}
\label{EfH}
\end{figure}

One of the main advantages of the hyperboloidal slicing method is the ability to normalize fields in their coordinates.  
The goal of this subsection is to provide a visualisation of that fact. 
Figure~\ref{Eftx} shows the eigenfunction corresponding to the fundamental mode as a function of the physical coordinate $x$. This eigenfunction diverges as $x \to \infty$. In contrast, Fig.~\ref{EfH} shows the same eigenfunction expressed in terms of the compactified coordinate $y$, where it becomes normalizable. This makes it possible to compute physically meaningful cavity quantities such as mode volume without ambiguity, regardless of the precise definition adopted.   
Note that the field representation differs between the original coordinates $(t,x)$ and the hyperboloidal coordinates $(\tau, y)$. 

\section{Spectral (in)-stability: Pseudospectrum}
\label{s:pseudospectrum}

Classical spectral theory establishes several key properties of self-adjoint operators: their eigenvalues are real, their eigenvectors form a complete orthonormal basis of the underlying Hilbert space, and their spectra are stable under small perturbations. These results extend naturally to normal operators (i.e. operators that commute with their adjoint), which may possess complex eigenvalues but still admit orthonormal eigenvectors and enjoy spectral stability. Formally, for a matrix $A \in M_n(\mathbb{C})$ with adjoint $A^*$, the operator is normal if and only if $[A,A^*] = 0$.  

By contrast, for non-normal operators no such comprehensive spectral theory exists. Their spectra may be highly sensitive to perturbations, and eigenvectors often fail to form a stable or complete basis. This instability is central to the analysis of quasi-normal modes in open optical systems.

\subsection{Standard notions of the pseudospectrum} 

Let $A$ be a linear operator. The spectrum $\sigma(A)$ is the set of complex numbers $\lambda$ for which the resolvent $R_A(\lambda) = (\lambda \mathrm{Id}-A)^{-1}$ fails to exist as a bounded operator.  

While this notion is well-defined for normal operators, it is insufficient for non-normal operators, where eigenvalues may exhibit extreme sensitivity. In such cases, studying only the spectrum provides an incomplete picture of operator behavior. A standard tool to capture this instability is the pseudospectrum.  

For $\epsilon > 0$, the $\epsilon$–pseudospectrum $\sigma_\epsilon(A)$ is defined equivalently by:
\bea
\label{e:pseudospectrum_def}
\sigma_\epsilon(A) &=& \{\lambda\in\mathbb{C}: \|(\lambda \mathrm{Id}- A)^{-1}\| > 1/\epsilon\} \nn \\
&=& \{\lambda\in\mathbb{C}: \exists v\in\mathbb{C}^n \;\; \|A v-\lambda v\| < \epsilon\}  \\
&=&  \{\lambda\in\mathbb{C}: \exists E\in M_n(\mathbb{C}), \|E\| < \epsilon,\; \lambda\in\sigma(A+E)\} \nn
\eea

Each formulation provides a distinct interpretation:
\begin{itemize}
\item The first definition characterizes pseudospectrum points as those where the resolvent norm is large. Unlike in the normal case, such points may lie far from the true eigenvalues.  
\item The second introduces the notion of an $\epsilon$–quasimode (or quasi-solution), fundamental in semiclassical analysis \cite{Sjostrand2019}. It also highlights the numerical challenge of computing spectra, since unavoidable round-off errors may produce spurious modes.  
\item The third shows that pseudospectrum points are exact eigenvalues of a perturbed operator with $\|E\| < \epsilon$, underscoring spectral instability under small perturbations \cite{JarMacAls22}.  
\end{itemize}

Pseudospectrum analysis was first introduced in gravitational-wave physics by \cite{Jaramillo:2020tuu} to quantify spectral instability, revealing subtle structures in black-hole perturbations. In the present work we adapt this framework to electromagnetic systems. Doing so opens the possibility of probing fine-scale features such as molecules, quantum dots, or minute changes in permittivity, and of predicting the spectral behavior of complex optical structures.

\subsection{Pseudospectrum for a generalized eigenvalue problem}
For our problem $L \psi = i \omega M \psi$, the pseudospectrum definitions generalize as follows:
\begin{align}
\sigma_\epsilon(L, M) 
&= \left\{ \lambda \in \mathbb{C} : \left\| (i\omega M - L)^{-1} \right\| > \frac{1}{\epsilon} \right\} \nonumber \\
&= \left\{ \lambda \in \mathbb{C} : \exists\, v \in \mathbb{C}^n,\; \| L v - \lambda M v \| < \epsilon \right\} \\
&= \left\{ \lambda \in \mathbb{C} : \exists\, E \in M_n(\mathbb{C}),\, \|E\| < \epsilon, \right. \nonumber \\
&\qquad\quad \left. \lambda \in \sigma(L + E, M) \right\}. \nonumber
\end{align}

\subsection{Numerical implementation}
A fourth equivalent characterization is particularly useful for computations:
\bea
\label{e:pseudospectrum_carac_G}
\sigma^\epsilon_{_E} (L, M) = \{\lambda\in\mathbb{C}: s_{_E}^{\min}(\lambda M - L) < \epsilon\},
\eea
where $s_{_E}^{\min}(Q) = \min \{\sqrt{\lambda}: \lambda \in \sigma(Q^\dagger Q) \}$ for $Q \in M_n(\mathbb{C})$.  
For implementation, the complex plane is discretized on a grid, and $s_{_E}^{\min}(i\omega M - L)$ is computed at each point. The $\epsilon$–pseudospectrum is then obtained as the set of points where this value is below $\epsilon$.
Numerical evaluation of the pseudospectrum thus requires computing singular values, which in turn depend on the choice of scalar product and the associated norm. In the following we contrast results obtained with the standard $L^2$ norm and with the energy norm, the latter providing a more physically meaningful perspective.
\subsection{Norms and QNM normalization}

The choice of the underlying norm is a fundamental aspect of pseudospectrum analysis. Since the $\varepsilon$-pseudospectrum is defined through perturbations measured by an operator norm, different norms may lead to different quantitative measures of spectral sensitivity and to distort pseudospectrum geometries. Consequently, the norm should be regarded as an intrinsic component of the stability analysis rather than merely as a numerical convention.

In optical physics, the notion of norm is closely related to the normalization of quasinormal modes (QNMs). Several normalization procedures have been proposed in the literature, among which the reciprocity-based normalization derived from the equal-frequency limit of Lorentz reciprocity is one of the most widely used~\cite{Muljarov:18,sauvan2013theory,LalYanVyn17}:
\begin{align}
	\mathcal{B}(\tilde{\phi}_m,\tilde{\phi}_n)
	&=
	\iiint_{\mathrm{space}}
	\left(
	\tilde{\mathbf{E}}_m \cdot
	\frac{\partial(\omega\varepsilon)}{\partial\omega}
	\tilde{\mathbf{E}}_n
	-
	\tilde{\mathbf{H}}_m \cdot
	\frac{\partial(\omega\mu)}{\partial\omega}
	\tilde{\mathbf{H}}_n
	\right)
	\,d^3\mathbf r .
\end{align}

From a mathematical perspective, this expression is more naturally interpreted as a reciprocity-based bilinear normalization form than as a Hilbert-space inner product. Although it naturally incorporates the dispersive properties of the medium through the frequency derivatives of the constitutive parameters, it is generally unconjugated and is not, in general, positive definite. Its primary purpose is therefore the normalization of resonant states and the determination of modal expansion coefficients rather than the definition of the norm underlying a pseudospectrum analysis.

In the present work, pseudospectra are computed with respect to the standard $L^2$ norm. This choice provides a mathematically well-defined reference that is independent of any particular modal normalization while allowing direct comparison with the existing pseudospectrum literature. Although the $L^2$ norm measures field amplitudes rather than electromagnetic energy, it constitutes a natural baseline for investigating spectral sensitivity.

It would nevertheless be of considerable interest to investigate pseudospectra associated with alternative physically motivated functional structures, including weighted energy norms and Sobolev norms~\cite{Gasperín_2022,Besson25,Jaramillo:2020tuu,JarMacAls22}. Such norms may provide a more physically meaningful measure of perturbation amplitudes and could reveal stability properties that remain hidden within the standard $L^2$ framework.

The distinction between modal normalization and the choice of norm is also reflected in modal expansions based on Keldysh's theorem. As discussed in~\cite{Besson25}, changing the normalization modifies the normalization constants and the modal expansion coefficients, while leaving the resonant expansion unchanged. By contrast, in pseudospectrum analysis, the choice of the underlying norm directly modifies the definition of perturbation size and therefore the quantitative notion of spectral sensitivity.

\subsection{Pseudospectrum results}
\subsubsection{Results in the $L^2$ norm}
\begin{figure}[H]
\centering
\includegraphics[scale=0.4]{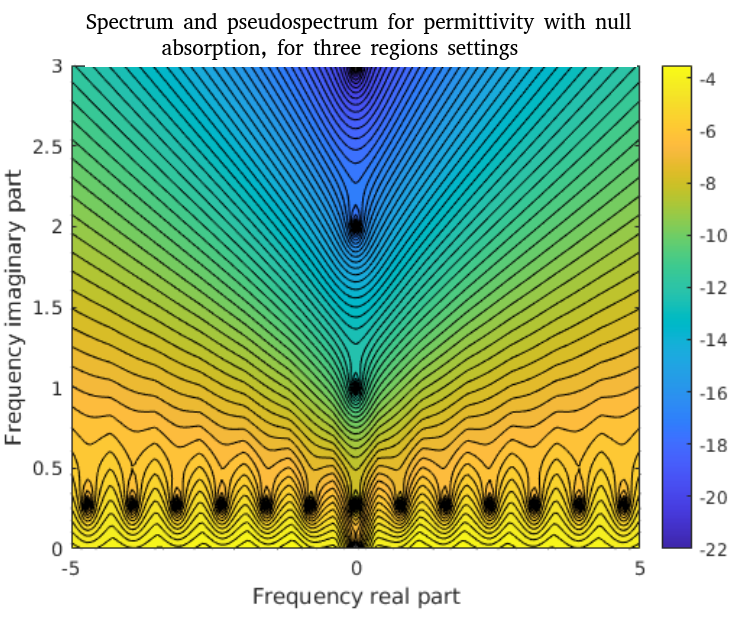}
\caption{Pseudospectrum in the $L^2$ norm for a non-dispersive cavity with constant permittivity $\epsilon_{II}=\text{cte}$ surrounded by air. 
Contour levels correspond to $\log_{10} \epsilon$, indicating the perturbation norm required to move eigenvalues. 
Circular pseudospectrum contours around each eigenvalue demonstrate stability against perturbations down to order $10^{-5}$, while eigenvalues with larger imaginary parts show greater sensitivity.}
\label{Per_1}
\end{figure}

\begin{figure}[H]
\centering
\includegraphics[scale=0.4]{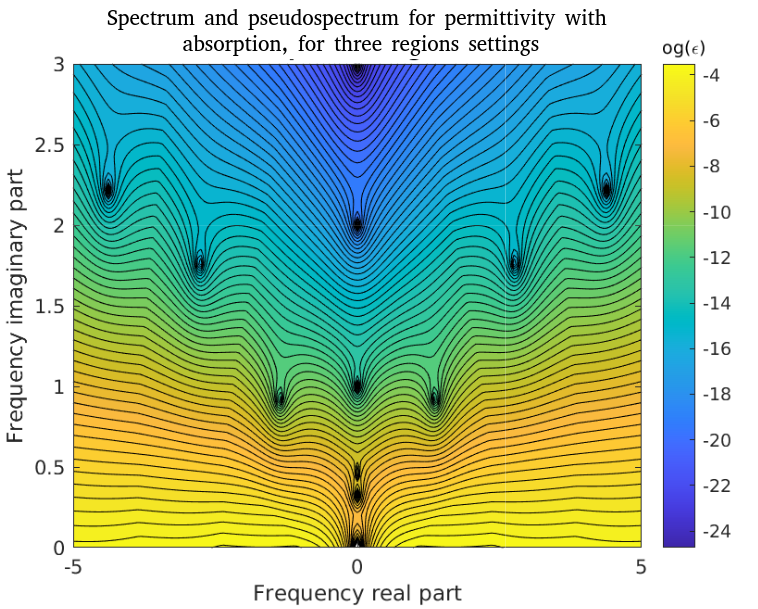}
\caption{Pseudospectrum in the $L^2$ norm for a dispersive cavity modeled by a Drude permittivity with $\omega_p=1$ and $\Gamma=0.1$. 
As in Fig.~\ref{Per_1}, concentric pseudospectrum contours are observed around the eigenvalues, but the distribution of eigenvalues differs due to dispersion and absorption. 
The comparison highlights how material dispersion modifies the sensitivity structure of the spectrum.}
\label{Per_2}
\end{figure}

Figures~\ref{Per_1} and \ref{Per_2} illustrate pseudospectra computed in the $L^2$ norm for two representative cases.  

Figure~\ref{Per_1} corresponds to a cavity with constant permittivity $\epsilon_{II}$. The color scale (logarithmic in $\epsilon$) indicates the perturbation norm required to displace eigenvalues. Around each eigenvalue, nearly concentric pseudospectrum contours appear, showing that perturbations of order $10^{-5}$ are sufficient to shift them significantly. Eigenvalues with larger imaginary parts are markedly more sensitive; for instance, the eigenvalue at $3i$ can be displaced by a perturbation of order $10^{-18}$.  

Figure~\ref{Per_2} presents the case of a dispersive, absorbing cavity described by the Drude model with $\omega_p=1$ and $\Gamma=0.1$. A similar pseudospectrum structure is observed, though the distribution of eigenvalues modifies the sensitivity pattern.  

These results naturally raise the question of their physical interpretation: what do the different perturbation orders signify? Following the reasoning of \cite{Jaramillo:2020tuu}, the energy norm provides a more physically meaningful assessment of stability, and is therefore considered next.

\subsection{Pseudospectrum in the energy norm (simplified case)}
The electromagnetic energy density is given by
\begin{equation}
	\label{EMe}
	\mathbf{W} = \frac{\epsilon_0 \mathbf{E}^2}{2} +\frac{(\frac{\partial \mathbf{P}}{\partial t})^2}{2 \omega_p^2 \epsilon_0} + \frac{\mu_0 \mathbf{H}^2}{2}.
\end{equation}

Since we are working in one dimension (normal incidence), the relevant nonzero field components reduce to $E_y(x)$ and $H_z(x)$, denoted hereafter as $E(x)$ and $H(x)$. Using Maxwell’s equations,
\begin{equation}
	\label{nabla}
	\begin{array}{lllll}
		\nabla \times \mathbf{E} = \frac{\partial E}{\partial x} \,\mathbf{\hat{z}}, \\
		\nabla \times \mathbf{H} = - \frac{\partial H}{\partial x} \,\mathbf{\hat{y}},
	\end{array}
\end{equation}
we obtain
\begin{equation}
	\label{maxwell3}
	\frac{\partial H}{\partial x} = - \epsilon_{\infty} \frac{\partial E}{\partial t} - \frac{\partial P}{\partial t}.
\end{equation}

By constructing an integration matrix from a Chebyshev expansion, $H$ can be computed from the left-hand side of \eqref{maxwell3}.  

For simplicity we set $\omega_0 = 0$, so that \eqref{polarization eq} reduces to
\begin{equation}
	\label{polarization eq2}
	\frac{\partial^2 P}{\partial t^2}+ \Gamma \frac{\partial P}{\partial t}=\omega_p^2 \epsilon_{\infty} E.
\end{equation}

The entire one-dimensional Maxwell system, coupled with a dispersive current, can now be written as:
\begin{equation}
	\begin{aligned}
		\partial_t H &= -\frac{1}{\mu_0}\,\partial_x E, \\
		\partial_t E &= -\frac{1}{\varepsilon_\infty}\,\partial_x H - \frac{1}{\varepsilon_\infty} J, \\
		\partial_t J &= -\Gamma J + \omega_p^2 \varepsilon_0 E.
	\end{aligned}
\end{equation}

\medskip
\noindent
Using auxiliary fields:
\begin{equation}
	\begin{aligned}
		A &= \partial_t E, \\
		J &= \partial_t P.
	\end{aligned}
\end{equation}
\medskip
\noindent

The equations can be arranged as follows:
\begin{align}
	\partial_t \big( \varepsilon_\infty A + J \big)
	&= \frac{1}{\mu_0} \partial_x^2 E, \\
	\partial_t E &= A, \\
	\partial_t J &= \omega_p^2 \varepsilon_0 E - \Gamma J.
\end{align}

\medskip

\noindent

Using the hyperboloidal transformation and proceeding as before, applying a Fourier transform,  this translates into the matricial form in \eqref{matricial form EM2}.  
\begin{equation}
	\label{matricial form EM2}
	\begin{aligned}
		\begin{bmatrix}
			L_1 & L_2 & 0 \\
			0 & I & 0 \\
			\omega_p^2 \varepsilon_\infty I & 0 & -\Gamma I
		\end{bmatrix} 
		\begin{bmatrix}
			E \\
			A \\
			J
		\end{bmatrix}
		&= \\
		i\omega\,
		\begin{bmatrix}
			0 & M_{12} & \mu_0 I \\
			I & 0 & 0 \\
			0 & 0 & I 
		\end{bmatrix} 
		\begin{bmatrix}
			E \\
			A \\
			J
		\end{bmatrix},
		&
	\end{aligned}
\end{equation}
where
\begin{equation}
	\begin{aligned}
		L_1 &= (1 - y^2)\left(-2y\,\partial_y + (1 - y^2)\,\partial_y^2 \right), \\
		L_2 &= (1 - y^2)\left(-2y\,\partial_y - I \right), \\
		M_{12} &= \varepsilon_\infty \mu_0\, I - y^2.
	\end{aligned}
\end{equation}

From this point, we follow the same steps to write a similar equation to \eqref{ev_prob_op}
and to impose the continuity conditions.
\textit{Note:} From Maxwell's equations,
\begin{equation}
	\partial_x H = -\varepsilon_\infty A - J,
\end{equation}
so that $H$ can be written in term of $A$ and $J$ after constructing the integration matrix.
\medskip
\noindent
\subsubsection{Results in the energy norm}
With the energy norm
\[
\| (E,P,H) \|^2 = \frac{\epsilon_0 E^2}{2} + \frac{(\partial_t P)^2}{2 \omega_p^2 \epsilon_0} + \frac{\mu_0 H^2}{2},
\]
we construct the adjoint operator and compute pseudospectra.  

\begin{figure}[H]
\centering
\includegraphics[scale=0.5]{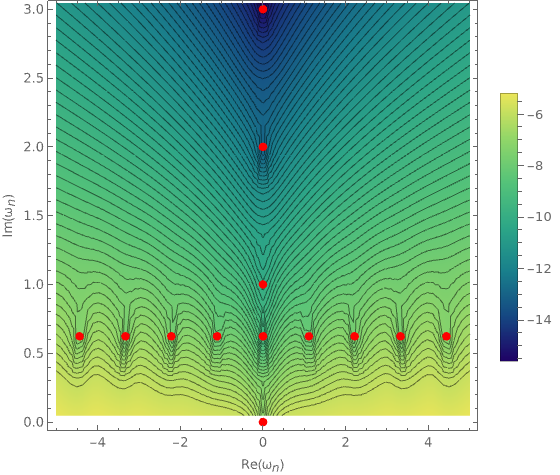}
\caption{Pseudospectrum in the energy norm for a non-dispersive cavity with constant permittivity $\epsilon_{II}=2$ and no absorption. 
Compared with the $L^2$ case (Figs.~\ref{Per_1}--\ref{Per_2}), the pseudospectrum scales differently with perturbation size, providing a physically meaningful measure of stability. 
Modes with the same imaginary part form pseudospectrum clusters, consistent with the expectation that only these contribute to the scattered-field expansion.}
\label{Per_EN-noabs}
\end{figure}

Figure~\ref{Per_EN-noabs} shows the pseudospectrum for a cavity with constant permittivity $\epsilon=2$. In this case the scattered field can be expressed as a sum over modes with equal imaginary parts, and the pseudospectrum retains a circular structure, though with different perturbation scales.  

\begin{figure}[H]
\centering
\includegraphics[scale=0.5]{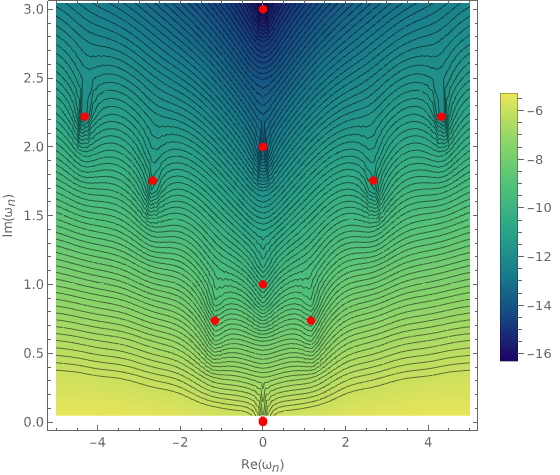}
\caption{Pseudospectrum in the energy norm for a dispersive cavity with Drude permittivity ($\omega_p=1$, $\Gamma=0.01$). 
Even eigenvalues with small imaginary parts become sensitive to perturbations of order $10^{-8}$, reflecting the strong influence of material absorption. 
This illustrates that the energy norm provides a sharper criterion for physical stability than the $L^2$ norm.}
\label{Per_EN}
\end{figure}

Figure~\ref{Per_EN} presents the dispersive case with Drude parameters $\omega_p=1$ and $\Gamma=0.01$. Even eigenvalues with small imaginary parts require perturbations no larger than $10^{-8}$ to be displaced.  

\begin{figure}[H]
\centering
\includegraphics[scale=0.5]{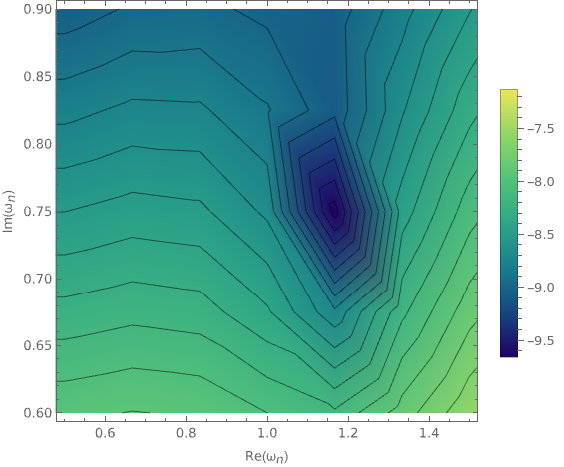}
\caption{Zoom-in of the energy-norm pseudospectrum around the eigenfrequency $\omega \approx 1.643 + 0.74i$ (taken from Fig.~\ref{Per_EN}). 
The local structure of the pseudospectrum reveals the anisotropic sensitivity of this mode: small perturbations move the eigenvalue preferentially along the imaginary axis. 
Such zoomed views highlight how pseudospectra capture mode-dependent stability characteristics.}
\label{Per_EN_p}
\end{figure}

Finally, Fig.~\ref{Per_EN_p} zooms in on the pseudospectrum near $\omega \approx 1.643 + 0.74i$, highlighting the local structure of the contours. An interesting open question concerns the behavior of purely imaginary eigenvalues under physical perturbations (e.g. variations in permittivity) and their potential impact on scattered-field expansions.

\subsection{Convergence of pseudospectrum}
\begin{figure}[H]
\centering
\includegraphics[scale=0.9]{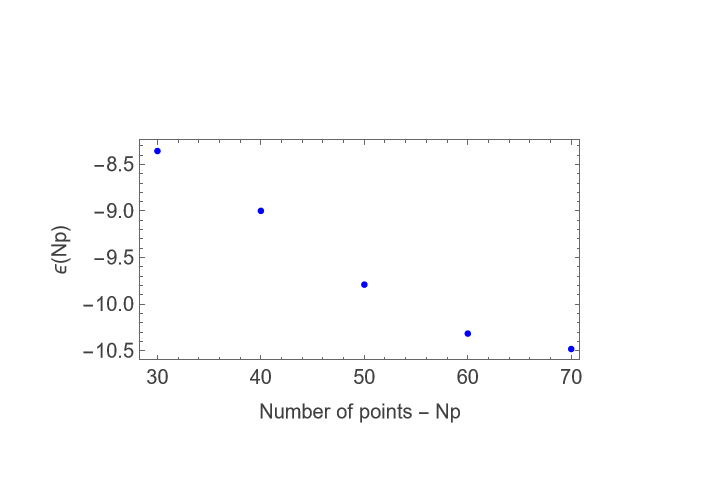}
\caption{Pseudospectrum convergence considering the point $1+2i$}
\label{ps_conv2}
\end{figure}

Let's test the convergence of the pseudospectrum (in the energy norm) for a random point in the lower part of the complex plane, let it be $1 +2 i$.
Considering a cavity with  the same parameters as in Fig.~\ref{Per_EN}  ($\omega_p=1$, $\Gamma=0.01$, $\omega_0=0$).
Fig.~\ref{ps_conv2} shows the order of perturbation as a function of increasing the number of points in the descretized grid.

\subsubsection{Physical possibilities of perturbation}

In the simplest cavity configurations, pseudospectrum analysis indicates susceptibility to instability. However, this does not by itself clarify how the eigenvalues behave under physically meaningful perturbations.  

\begin{figure}[H]
\centering
\includegraphics[scale=0.6]{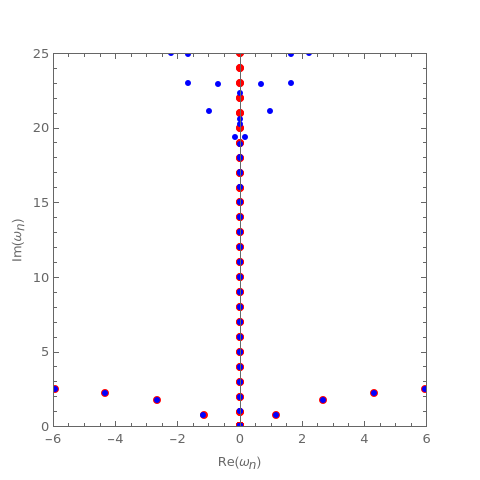}
\caption{Spectrum under a perturbation of order $10^{-6}$ applied to the cavity permittivity. 
Red dots indicate eigenvalues of the unperturbed operator; blue dots correspond to the perturbed operator. 
Eigenvalues on the logarithmic Regge branches remain essentially stable, while purely imaginary modes are shifted and acquire real parts. 
This confirms the prediction from the pseudospectrum analysis that anti-bound states are the most sensitive to physical perturbations.}
\label{Per_6}
\end{figure}

To illustrate this point, Fig.~\ref{Per_6} shows the spectrum of a cavity with the same parameters as in the previous section, perturbed by modifying the permittivity at order $10^{-6}$.

It is apparent that the eigenvalues on the logarithmic branches remain stable under this perturbation, while the purely imaginary eigenvalues begin to split.  

\textit{\textbf{Interpretation}}  
\begin{itemize}
\item Eigenvalues located on the ``logarithmic'' branches appear stable. These modes lie on contour lines that delimit regions of the $\epsilon$–pseudospectrum. Following the results of \cite{Jaramillo:2020tuu}, \cite{JarMacAls22}, \cite{Alsheikh21}, eigenvalues tend to move along pseudospectrum contour lines under perturbation. This does not guarantee immobility, but it does constrain their displacement. Further studies are needed to fully characterize this behavior.  
\item Eigenvalues with larger imaginary parts are the first to be affected by perturbations, consistent with the gravitational-wave case where similar pseudospectrum structures arise. Once shifted, these eigenvalues acquire a nonzero real part, indicating that the associated modes transition from purely decaying to oscillatory-decaying behavior. This effect deserves closer investigation.  
\end{itemize}

What has been demonstrated so far is a framework for analyzing stability and instability, providing a foundation for further physical interpretation. For realistic studies, however, additional refinements are required, including the incorporation of Kramers--Kronig relations to ensure that the permittivity satisfies causality. These relations connect the real and imaginary parts of permittivity and thus impose essential physical constraints.

\subsection{Weyl law}

Weyl’s law describes the asymptotic number of eigenvalues of the Laplace–Beltrami operator (Eq.~\ref{e:Laplace_Beltrami}) on a compact domain $D$, subject to homogeneous boundary conditions, in the high-frequency limit:
\bea
\label{e:Laplace_Beltrami}
\Delta f = \frac{1}{\sqrt{|g|}} \partial_i \left(\sqrt{|g|} g^{ij} \partial_j f \right), 
\eea
where $g_{ij}$ is the metric tensor.  

Let $\lambda_n$ denote the eigenvalues of $\Delta$. Then the counting function $N(\lambda)$, i.e. the number of eigenvalues with $|\lambda_n| < \lambda$, satisfies
\bea
\label{e:Weyl_LB}
N(\lambda) \sim \mathrm{Vol}_d(D)\, \lambda^{\frac{d}{2}} +  o(\lambda^{\frac{d-1}{2}}), \quad (\lambda \to \infty),
\eea
where $\mathrm{Vol}_d(D)$ is the volume of the compact domain and $d = \dim(D)$.  

Similar asymptotics hold for other selfadjoint operators \cite{Ivrii16}, where the spectral theorem ensures discreteness and completeness of the modes. In \cite{Jar24}, the authors conjectured that a version of Weyl’s law also applies to non-selfadjoint black-hole operators.  

Here we suggest that open optical cavities, where the scattering structure is encoded in the permittivity, likewise obey a Weyl-type law. In this context, $\lambda$ of the Laplace–Beltrami operator corresponds to $\omega^2$, so the law implies scaling proportional to $(\omega^2)^{d/2}$.  

\begin{figure}[H]
\centering
\includegraphics[scale=1]{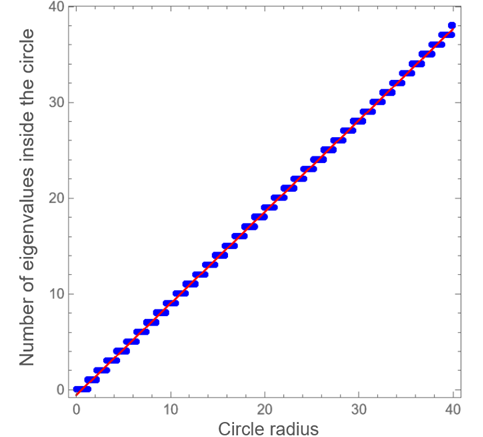}
\caption{Verification of Weyl-type scaling for eigenfrequencies for a cavity with $\omega_p=1$, $\Gamma=0$, $\omega_0=0$. 
The blue staircase function gives the number of eigenfrequencies inside a circle as a function of a circle radius in the complex plane. The red line is the identity line.   
The nearly linear scaling confirms the validity of a Weyl law analogue in this optical setting, with the proportionality length equal to the cavity length.}
\label{Weyl1}
\end{figure}

Numerical computations for the case $\omega_p=1$, $\Gamma=0$, $\omega_0=0$ show that the Weyl length coincides with the physical cavity length. Figure~\ref{Weyl1} demonstrates the proportionality between the number of eigenfrequencies inside a circle in the complex plane and the circle’s radius. Notably, in this case the Weyl law holds not only asymptotically but across the full spectral range.

\section{Conclusions and Perspectives}

\subsection{Conclusions}

We have presented a novel spectral and geometric framework for analyzing quasi-normal modes (QNMs) in open optical cavities. This framework combines the \textit{hyperboloidal slicing approach}—originally developed for gravitational wave and more generally gravity physics~\cite{Ansorg_2016,Macedo_2018,Macedo_2020,Zengino_lu_2011,Bizon:2014nla,Donninger:2020sqm,MacZen25}—with \textit{pseudospectrum analysis}~\cite{Jaramillo:2020tuu,JarMacAls22,Alsheikh21} to investigate spectral stability and sensitivity in electromagnetic systems.

By reformulating the Maxwell–Lorentz equations in compactified hyperboloidal coordinates and introducing auxiliary field variables, we have posed the QNM eigenvalue problem in a Hilbert space without the need for artificial absorbing boundaries such as perfectly matched layers (PML). The use of \textit{Chebyshev spectral methods}~\cite{Ansorg:2016ztf,canuto2007spectral,trefethen2000spectral} provides highly accurate computations of complex eigenfrequencies for non-selfadjoint operators.

Two particularly significant conclusions emerge:

\begin{enumerate}
    \item \textbf{Purely imaginary eigenvalues.}  
    Our computations reveal a discrete set of purely imaginary eigenfrequencies (“anti-bound” states), previously reported in other contexts~\cite{HokNob65,OhaGin74,Belchev2011,Siegert39,Gamow28,Peierls59} but not yet fully understood physically. In our framework, these modes are \textit{sensitive to physical perturbations} of the permittivity, which can induce nonzero real parts in their frequencies, thereby altering their dynamical role. This sensitivity suggests they warrant further targeted study in both frequency and time domains.
    
    \item \textbf{Norm dependence of the pseudospectrum.}  
    While the $L^2$ pseudospectrum (Figs.~8--9) reveals the general structure of spectral sensitivity, the \textit{energy norm} pseudospectrum (Figs.~10--12) provides a more physically meaningful picture of instability under realistic perturbations. This not only highlights the importance of carefully choosing the scalar product when assessing resonance robustness, but also demonstrates the presence of potential spectral instabilities in optical systems.
\end{enumerate}

Additional findings include:

\begin{itemize}
    \item Normalization in compactified coordinates enables rigorous computation of cavity characteristics such as mode volume (Figs.~4--7).
    \item Physical perturbations of the permittivity (Fig.~13) show that eigenvalues on Regge-type QNM logarithmic branches remain relatively stable, whereas purely imaginary modes are more vulnerable.
    \item The Weyl-type scaling observed in eigenvalue counting (Fig.~14) suggests that an adapted Weyl law may hold for non-selfadjoint electromagnetic operators, in analogy with conjectures in gravitational systems~\cite{Jar24}.
\end{itemize}

Together, these results establish a methodological bridge between gravitational and optical QNM theory, offering a new quantitative foundation for \textit{stability assessment, sensitivity analysis, and modal control} in photonic resonators.

\subsection{Perspectives}

Building on this framework, several avenues of future research are suggested:

\begin{enumerate}
    \item \textbf{Interpretation of purely imaginary eigenvalues.}  
    Clarify their physical meaning and dynamical role, especially in light of their appearance in other fields~\cite{HokNob65,OhaGin74,Belchev2011,Siegert39,Gamow28,Peierls59} and their observed sensitivity to realistic perturbations. Investigate their contribution to the scattered field in both frequency and time domains.
    
    \item \textbf{Realistic material models.}  
    Extend the analysis to dispersive and absorptive media that satisfy \textit{Kramers–Kronig relations}~\cite{Muljarov:18,sauvan2013theory,LalYanVyn17}, ensuring causal consistency and enabling comparison with experimental data.
    
    \item \textbf{Physically motivated perturbations.}  
    Study geometric, material, and environmental perturbations to assess stability in realistic scenarios, building on sensitivity analyses developed in gravitational QNMs~\cite{Jaramillo:2020tuu,JarMacAls22}.
    
    \item \textbf{Time-domain implications and transients.}  
    Link pseudospectrum structures—especially in the energy norm—to \textit{transient amplification} phenomena~\cite{Gasperín_2022,Besson25}, exploring how norm choice influences predictions.
    
    \item \textbf{Higher-order norms and Sobolev analysis.}  
    Investigate pseudospectra in Sobolev $H^p$ norms~\cite{Jaramillo:2020tuu,Besson25,Gasperín_2022} to uncover hidden instability mechanisms tied to higher-order derivatives.
    
    \item \textbf{Generalization to higher dimensions.}  
    Extend the combined hyperboloidal and pseudospectrum framework to two- and three-dimensional resonators, where geometry-driven non-normality could generate richer instability patterns.
    
    \item \textbf{Weyl law extensions.}  
    Test the applicability of Weyl-type eigenvalue counting in non-selfadjoint optical cavities and explore deviations caused by openness, dispersion and geometry.
\end{enumerate}

By combining \textit{geometric hyperboloidal compactification} with \textit{pseudospectrum sensitivity analysis in physically relevant norms}, this study introduces a robust computational toolset and points toward a richer understanding of \textit{modal (in)stability and resonance behavior} across diverse wave-physics research domains. Beyond optics, the framework can be adapted to other problems involving wave propagation with outgoing boundary conditions, e.g. \textit{heat transfer, acoustics, condensed matter or physical oceanography, among others}.

\section{Acknowledgments}

The authors would like to warmly thank Rodrigo P.~Macedo for the incredibly stimulating discussions and to Yann Boucher and Brian Stout for their generous sharing of ideas, We are also grateful to Marcus Ansorg and his work on spectral methods.
We would like to also thank Johannes Sj$\ddot{o}$strand, Michael Hitrik and Giuseppe Dito for the enriching insights on spectral theory and functional analysis.
Many thanks to Hans Jauslin, G\'erard Colas de Frank, Philippe Lalanne and K\'evin Vynck for the interesting exchange and perspectives. 

Thanks are extended to J\'er\'emy Besson, Edgar Gasper\'in, Oscar Meneses-Rojas and Mokdad Mokdad for helpful and interesting discussions.

This work was supported by the PO FEDER-FSE Bourgogne 2014/2020 program, the EIPHI Graduate School (contract ANR-17-EURE-0002), and the "Investissements d’Avenir" program through project ISITE-BFC (ANR-15-IDEX-03). Additional support came from the ANR 'Quantum Fields interacting with Geometry' (QFG) project (ANR-20-CE40-0018-02).

\appendix
\section{Reduced Drude Model}
\subsection{Permittivity of generic case following Lorentz model}
Making an educated assumption that the polarization is exponentially dependent on time:
$P(t)=P_0 e^{-i \omega t }$ and inseting that in eq.\ref{polarization eq}, we get the general form of a permittivity that follows Lorentz model:
\begin{equation}
(- \omega^2 -i \Gamma \omega + w_0^2) \tilde{P} = \omega_p^2 \epsilon_\infty E_0
\end{equation}
and then:
\begin{equation}
\label{EP_relation}
\frac{\tilde{P}}{E_0}=\frac{\omega_p^2 \epsilon_\infty}{-\omega^2 -i \Gamma \omega + w_0^2}
\end{equation}
On the other hand the relation between the electric field and the polarization is:
\begin{equation}
P= \epsilon_0 \chi_e E_0,
\end{equation} 
where $\chi_e$ is the electric susceptibility of the medium and it is related to its relative permittivity by the relation: $\chi_e =\epsilon_\infty (\epsilon_r -1)$
plugging the last two equations in \ref{EP_relation} one gets:
\begin{equation}
\epsilon_r = \epsilon_\infty - \frac{\omega_p^2}{\omega^2+i \Gamma \omega - \omega-0^2 }
\end{equation}
Assuming $\omega_0=0$ so we are studying a metal case, one gets Drude model for metals:
\begin{equation}
\label{Drude_M}
\epsilon_r = \epsilon_\infty - \frac{\omega_p^2}{\omega^2+i \Gamma \omega  }
\end{equation}

\subsection{Contact with a scattering by a potential}
Putting $\Gamma$ to $0$ (zero absorption) and plugging that in the QNMs wave equation we get:
\begin{equation}
\Delta E + \frac{\epsilon_\infty}{c^2} E - w_p^2 E=0
\end{equation}
Which is exactly the QNMs wave equation with a potential $V= w_p^2$.
Having that makes a potential case a special "mathematical" case of the permittivity one. This opens a door to exchange knowledge and some results between two major fields gravitational and optics.

\section{Chebyshev collocation grids}
\label{ChebyApp}
In this appendix we provide technical information and equations which were used to implement the methods we used.

Spectral methods are widely used in applied mathematics for solving differential and eigenvalue problems \cite{canuto2007spectral,trefethen2000spectral,trefethen2005spectra}.  

Several numerical approaches exist for handling partial differential equations, including finite difference methods (FDM), finite element methods (FEM), and spectral methods. All are based on discretizing the domain $I$ and seeking an approximate solution that minimizes the residual error. In these approaches, the solution (trial function) is expanded in terms of basis functions (approximating functions), while test functions are used to minimize the residual generated by substituting the trial function into the governing equations.  

The finite difference method is the simplest, but provides only pointwise approximations. The finite element method is widely used in physics and engineering due to its flexibility in handling complex geometries, dividing the domain into subdomains with local, piecewise-defined approximating functions that are not globally smooth. Spectral methods, in contrast, provide global expansions of the solution and are known for their efficiency and accuracy in solving ordinary and partial differential equations, as well as eigenvalue problems. Because they are global, the derivative of a function at one point depends on the function values across the entire domain, and the resulting solutions are globally smooth.  

There are three principal variants of spectral methods—collocation, Galerkin, and tau—depending on the choice of test functions. The tau method enforces orthogonality of the residual against as many basis functions as possible, while the Galerkin method enforces it against recombined bases. For non-selfadjoint operators, particular care must be taken in choosing the scalar product. The collocation method avoids this difficulty by choosing the test functions as delta functions at selected collocation points, requiring that the residual vanish exactly at those points.  

The collocation method was first developed for spatially periodic problems by Kreiss and Oliger (1972) \cite{kreiss1972comparison} and by Orszag (1972) \cite{orszag1972comparison}. Orszag’s earlier work (1969) \cite{orszag1969numerical} laid the foundations for spectral methods in general. Subsequent studies comparing different numerical approaches \cite{orszag1972comparison,6,7,nasa83,8} consistently found collocation to be at least as efficient, and often the most efficient, method.  

For non-periodic systems, algebraic polynomial bases such as Chebyshev polynomials are required.
\subsection{Chebyshev polynomials}

Chebyshev polynomials of the first kind are defined through the identity :
\begin{equation}
T_k(cos \theta) =\cos(k \theta)
\end{equation}
It can be written also for $|x| < 1$ as: 
\begin{equation}
T_k(x) =\cos(k \:  arccos(x))
\end{equation}
The following lines gives the few first Chebyshev polynomials:

\begin{equation}
\begin{split}
&T_0(x) = 1 \\
& T_1(x) = x \\
&T_2(x) = 2x^2-1 \\
&T_3(x) = 4x^3-x
\end{split}
\end{equation}
Here are a few of their important properties, which will help for the rest of the chapter:\\
\begin{itemize}
\item \textbf{Recursion relation}\\
Because of:$\cos((k+1)\phi) +\cos((k-1) \phi) = 2\cos(k \phi)\cos(\phi)  $
Then Chebyshev polynomials satisfy the following recursion relation:
\begin{equation}
T_{k+1} = 2xT_k(x) - T_{k-1}(x)
\end{equation}
\item \textbf{Orthoganality relation} \\
Chebyshev polynomials are orthogonal with respect to the weighting function $\frac{1}{\sqrt(1-x^2)}$, on the interval $[-1,-1]$  they verify:
\begin{align}
\label{Chebyshev_orthogonality}
\langle T_n, T_m \rangle_w
&= \int_{-1}^{1} \frac{T_n(x)\, T_m(x)}{\sqrt{1-x^2}}\, dx \\[1.5ex]
&= \int_0^{\pi} \cos(n\phi)\, \cos(m\phi)\, d\phi \notag \\[1.5ex]
&=
\begin{cases}
      0 & m \neq n \\[1ex]
      \dfrac{\pi}{2} & m = n \neq 0 \\[2ex]
      \pi & m = n = 0
\end{cases}
\notag
\end{align}

\item \textbf{Expansion} \\
Let $ \psi (x) \in \mathbb{L}^2_{w}([-1,+1]) $, then it can be written as:
\begin{equation}
\label{exp1}
\psi(x) = \frac{c_0^{\psi}}{2} + \sum_1^{\infty} c_k^{\psi} T_k(x)
\end{equation}
\end{itemize}

\subsection{Grids}
\begin{itemize}
\item Chebyshev-Gauss grid, where the points corresponds to the roots of Chebyshev polynomial.
\begin{equation}
\label{Gauss_points}
x_j =\cos(\frac{\pi (j+\frac{1}{2})}{N+1}): j=0,1,2,....,N
\end{equation}
And $\phi_j=\frac{\pi (j+\frac{1}{2})}{N+1}$.
Note that this grid contains no boundary points, $x_0 =\cos(\frac{\pi}{2N+2}) < 1$, and
 $x_N = -\cos(\frac{\pi}{2N+2}) > -1$.
 
\item Chebyshev-Lobatto grid, where the points are the extrema of Chebyshev polynomial:
\begin{equation}
\label{Lobatto_points}
x_j=cos(\frac{j \pi}{N} ): j=0,1,2,...,N
\end{equation}
Note that this grids contain the two boundary points, $x_0 = +1$, and $x_N = -1$.
\item Right Chebyshev-Radau grid. Its points correspond to the following formula:
\begin{equation}
\label{Left_Radaup}
x_j =\cos(\frac{2\pi j}{2N+1}): j=0,1,2,....N
\end{equation}
This corresponds to the Fourier points: $\phi_j= \frac{2 \pi j}{2N+1}$.
So this grid contains the right boundary edge $x_0 = +1$ but not the left one $x_0 =\cos(\frac{2 \pi N}{2N+1})> -1$.

\item Left Chebyshev-Radau grid. Its points correspond to the following formula:
\begin{equation}
\label{Left_Radau}
x_j = -\cos\left( \frac{2\pi j}{2N+1} \right)
= \cos\left( \frac{ \pi(2N+1-2j)}{2N+1} \right),
\quad j = 0,1,2,\ldots,N
\end{equation}

This corresponds to the Fourier points: $\phi_j=\pi - \frac{2 \pi j}{2N+1}$.
So this grid contains the left boundary edge $x_0 = -1$ but not the right one $x_0 =\cos(\frac{\pi}{2N+1}) < 1$.
\end{itemize}

\subsection{Chebyshev coefficients}
Chebyshev expansion approximation:
\begin{equation}
\label{exp2}
{\psi}(x) = \frac{c_0}{2} + \sum_1^{N} c_k T_k(x)
\end{equation}
\begin{itemize}
\item Coefficients in Gauss grid: 
\begin{equation}
\label{Gauss_coefficients}
c_m = \frac{2}{N+1}  \sum_{j=0}^{N} \psi(x_j) T_m(x_j)
\end{equation}
\item Coefficients in Lobatto grid:

\begin{equation}
\label{Lobatto_coefficients}
c_m = \frac{2-\delta_{mN}}{2N} (\psi(1) + (-1)^m \psi(-1) + 2 \sum_{j=1}^{N-1} \psi(x_j) T_m(x_j))
\end{equation}
\item Coefficients in Right-Radau grid:
\begin{equation}
\label{RRadau_coefficients}
c_m = \frac{4}{2N+1} (\frac{\psi(1)}{2} + \sum_{j=1}^{N} \psi(x_j) T_m(x_j))
\end{equation}
\item Coefficients in Left-Radau grid:
\begin{equation}
c_m = \frac{4}{2N+1} (\frac{(-1)^m \: \psi(-1)}{2} + \sum_{j=1}^{N} \psi({x_j}) T_m(x_j))
\end{equation}
\end{itemize}
\subsection{Chebyshev differential matrix}

\textit{Gauss}\\
\begin{equation}
D_{mj}^1 =
\begin{cases}
\frac{x_m}{2(1-x_m^2)} & m=j \\[3ex]
(-1)^{m-j} \frac{\sqrt{1-x_j^2}}{(x_m - x_j) \sqrt{1-x_m^2}} &  m \neq j
\end{cases}
\end{equation}
and those of the second order  differentiation matrix:
\begin{equation}
D_{mj}^2 =
\begin{cases}
\frac{x_m}{(1-x_m^2)^2} - \frac{N(N+2)}{3(1-x_m^2)}& m=j \\[3ex]
(-1)^{m-j} \frac{\sqrt{1-x_j^2}}{(x_m - x_j) \sqrt{1-x_m^2}} (\frac{x_m}{(1-x_m^2)} - \frac{2}{x_m-x_j}) &  m \neq j
\end{cases}
\end{equation}

\textit{Lobatto}
\begin{equation}
D_{mj}^1 =
\begin{cases}
-\frac{2N^2+1}{6} (N+1) & m=j=0 \\[3ex]
\frac{2N^2+1}{6} (N+1) & m=j=N \\[3ex]
\frac{-x_j }{2(1-x_j)^2} & m=j \neq 0,N \\[3ex]
\frac{\kappa_m}{\kappa_j}\frac{(-1)^{m-j}}{(x_m-x_j)}  & m \neq j
\end{cases}
\end{equation}
and those of the second order  differentiation matrix:
\begin{equation}
D_{mj}^2 =
\begin{cases}
\frac{N^2 -1}{15}  & m=j=0,N
\\[3ex]
 \frac{-1}{(1-x_j^2)^2} - \frac{N^2-1}{3(1-x_j^2)} &  m=j \neq 0,N 
 \\[3ex]
\frac{2} {3} \frac{(-1)^j}{\kappa_j} \frac{(2N^2+1)(1-x_j)-6}{(1-x_j)^2}  & 0 = m \neq j \\[3ex]
\frac{2} {3} \frac{(-1)^{N+j}}{\kappa_j} \frac{(2N^2+1)(1+x_j)-6}{(1+x_j)^2}  & N = m \neq j \\[3ex]
\frac{(-1)^{m-j}}{\kappa_j} \frac{x_m^2+x_m x_j-2}{(x_m -x_j)^2 (1-x_m^2)} & 0 \neq m \neq N ,  j \neq m
\end{cases}
\end{equation}
where:
\begin{equation}
\kappa_j= 
\begin{cases}
1 & 0 < j < N \\
2 & j=0,N
\end{cases}
\end{equation}
\textit{Right Radau}\\
\begin{equation}
D_{mj}^1 =
\begin{cases}
\frac{N}{3} (N+1) & m=j=0 \\[3ex]
(-1)^{j} \frac{2 \sqrt{1+x_j} }{(1-x_j)} & m=0, j \neq 0 \\[3ex]
\frac{(-1)^{m+1}}{\sqrt{2} (1-x_m^2) \sqrt{1+x_m}}  & m \neq 0, j=0 \\[3ex]
\frac{-1}{2 (1-x_m^2)} & m=j \neq 0 \\[3ex]
\frac{(-1)^{m-j}}{x_m -x_j} \sqrt{\frac{1+x_j}{1+x_m}} & 0 \neq m \neq j \neq 0
\end{cases}
\end{equation}

and those of the second order  differentiation matrix:

\begin{equation}
D_{mj}^2 =
\begin{cases}
\frac{1}{15} (N-1)N(N+1)(N+2) & m=j=0 
\\[2ex]
(-1)^{j} \frac{2\sqrt{2} \sqrt{1+x_j} }{3(1-x_j)^2} [N (N+1)(1-x_j)-3] & m=0, j \neq 0
 \\[2ex]
\frac{(-1)^{m+1} (2x_m+1)}{\sqrt{2} (1-x_m^2)^2 (1+x_m)^{\frac{3}{2}}}  & m \neq 0, j=0 
\\[2ex]
\frac{-N(N+1)}{3 (1-x_m^2)} - \frac{x_m}{ (1-x_m^2)^2} & m=j \neq 0 
\\[2ex]
\frac{(-1)^{m-j}(2x_m^2-x_m+x_j-2)}{(x_m -x_j)^2 (1-x_m^2)} \sqrt{\frac{1+x_j}{1+x_m}} & 0 \neq m \neq j \neq 0
\end{cases}
\end{equation}

\newpage
\textit{Left Radau}\\
\begin{equation}
D_{mj}^1 =
\begin{cases}
\frac{-N}{3} (N+1) & m=j=0 \\[3ex]
(-1)^{j+1} \frac{2 \sqrt{1+x_j} }{(1-x_j)} & m=0, j \neq 0 \\[3ex]
\frac{(-1)^{m}}{\sqrt{2} (1-x_m^2) \sqrt{1+x_m}}  & m \neq 0, j=0 \\[3ex]
\frac{1}{2 (1-x_m^2)} & m=j \neq 0 \\[3ex]
\frac{(-1)^{m-j}}{x_j - x_m} \sqrt{\frac{1+x_j}{1+x_m}} & 0 \neq m \neq j \neq 0
\end{cases}
\end{equation}

and those of the second order  differentiation matrix:

\begin{equation}
D_{mj}^2 =
\begin{cases}
\frac{1}{15} (N-1)N(N+1)(N+2), & m=j=0, \\[3ex]
(-1)^{j} \frac{2\sqrt{2} \sqrt{1+x_j} }{3(1-x_j)^2} \left[ N (N+1)(1-x_j)-3 \right], & m=0, \ j \neq 0, \\[3ex]
\frac{(-1)^{m+1} (2x_m+1)}{\sqrt{2} (1-x_m^2)^2 (1+x_m)^{\frac{3}{2}}}, & m \neq 0, \ j=0, \\[3ex]
\frac{-N(N+1)}{3 (1-x_m^2)} - \frac{x_m}{ (1-x_m^2)^2}, & m=j \neq 0, \\[3ex]
\frac{(-1)^{m-j}(2x_m^2-x_m+x_j-2)}{(x_m -x_j)^2 (1-x_m^2)} 
\sqrt{\frac{1+x_j}{1+x_m}}, & m \neq 0, \ m \neq j, \ j \neq 0.
\end{cases}
\end{equation}
\newpage

\subsection{Chebyshev integration matrix}
To integrate a function over the space in a Lobatto grid, we can use the matrix
\begin{align*}
I_{ij} =
\begin{cases}
\begin{aligned}
&\dfrac{\cos\left(\frac{\pi i}{N_z}\right)}{2 N_z}
+ \dfrac{1 + \cos\left(\frac{2\pi i}{N_z}\right)}{4 N_z} \\[1ex]
&\quad + \displaystyle\sum_{k=2}^{N} 
\dfrac{2 - \delta_{k,N}}{2 N} \;
\dfrac{
    \cos\left(\frac{\pi i k}{N_z}\right)\cos\left(\frac{\pi i}{N_z}\right)
    + k \sin\left(\frac{\pi i k}{N_z}\right)\sin\left(\frac{\pi i}{N_z}\right)
}{
    1 - k^2
}
\end{aligned}
& j=0 \\[4ex]
\begin{aligned}
&\dfrac{\cos\left(\frac{\pi i}{N_z}\right)}{2 N_z}
- \dfrac{1 + \cos\left(\frac{2\pi i}{N_z}\right)}{4 N_z} \\[1ex]
&\quad + \displaystyle\sum_{k=2}^{N} 
(-1)^k\, \dfrac{2 - \delta_{k,N}}{2 N} \;
\dfrac{
    \cos\left(\frac{\pi i k}{N_z}\right)\cos\left(\frac{\pi i}{N_z}\right)
    + k \sin\left(\frac{\pi i k}{N_z}\right)\sin\left(\frac{\pi i}{N_z}\right)
}{
    1 - k^2
}
\end{aligned}
& j=N \\[4ex]
\begin{aligned}
&\dfrac{\cos\left(\frac{\pi i}{N_z}\right)}{N_z}
- \dfrac{1 + T_1(x(j+1))\cos\left(\frac{2\pi i}{N_z}\right)}{2 N_z} \\[1ex]
&\quad + \displaystyle\sum_{k=2}^{N}
\dfrac{2\,(2 - \delta_{k,N})}{2 N} \;
T_k\big(x(j+1)\big)\,
\dfrac{
    \cos\left(\frac{\pi i k}{N_z}\right)\cos\left(\frac{\pi i}{N_z}\right)
    + k \sin\left(\frac{\pi i k}{N_z}\right)\sin\left(\frac{\pi i}{N_z}\right)
}{
    1 - k^2
}
\end{aligned}
& 1 \leq j \leq N-1
\end{cases}
\end{align*}

\section{Derivation details in hyperboloidal coordinates}

For completeness, we include here the detailed derivations of the operator transformations used in the main text.  

\subsection{Coordinate transformation}
Starting from the Bizo\'n--Mach hyperboloidal coordinates
\begin{equation}
\tau = t - \ln(2\cosh x), \qquad y = \tanh(x),
\end{equation}
the tangent vectors transform according to
\begin{equation}
\begin{aligned}
\partial_t &= \partial_\tau, \\
\partial_x &= - y \, \partial_\tau + (1-y^2) \, \partial_y.
\end{aligned}
\end{equation}
\vspace{11cm}
\subsection{Second-order derivatives}
From these relations, one computes
\begin{equation}
\begin{aligned}
\partial_t^2 &= \partial_\tau^2, \\
\partial_x^2 &= y^2 \, \partial_\tau^2 - 2y(1-y^2) \, \partial_\tau \partial_y 
- (1-y^2)\,\partial_\tau \\
&\quad - 2y(1-y^2)\,\partial_y + (1-y^2)\,\partial_y^2.
\end{aligned}
\end{equation}

\subsection{Differential operators $L_1$ and $L_2$}
Inserting these expressions into the Maxwell--Lorentz system yields the operators
\begin{equation}
 \begin{aligned}
 L_1 &= - 2y (1-y^2) \partial_y + (1-y^2)^2 \partial_y^2, \\
 L_2 &= - 2y (1-y^2) \partial_y - (1-y^2) I,
 \end{aligned}
\end{equation}
as given in Eq.~(\ref{matricial form EM}) of the main text.

\subsection{Scalar reduction}
In the one-dimensional reduction with fields aligned along a fixed axis, 
the vector fields $\mathbf{E}$ and $\mathbf{P}$ reduce to scalar fields $E$ and $P$. 
This simplification transforms Maxwell’s curl equations into one-dimensional derivatives while retaining the dispersive structure.

\bigskip
These steps justify the compact operator form presented in the main body of the article.

\bibliographystyle{spmpsci}
%

\bibliography{Biblio_Op}

\end{document}